 \documentclass{emulateapj}

\usepackage{color}

\shorttitle{The ISM of two interacting distant starbursts at high resolution}
\shortauthors{Oteo et al.}

\usepackage{natbib}

\begin{document}


\title{Witnessing the birth of the red sequence: ALMA high-resolution imaging of [CII] and dust in two interacting ultra-red starbursts at $z = 4.425$}


\author{I. Oteo\altaffilmark{1,2},
R. J. Ivison\altaffilmark{2,1},
L. Dunne\altaffilmark{1,3}, 
I. Smail\altaffilmark{4},
M. Swinbank\altaffilmark{4,5},
Z-Y. Zhang\altaffilmark{1,2},
A. Lewis\altaffilmark{1},
S. Maddox\altaffilmark{1,3},
D. Riechers\altaffilmark{6},
S. Serjeant\altaffilmark{7},
P. Van der Werf\altaffilmark{8},
M. Bremer\altaffilmark{9},
P. Cigan\altaffilmark{3},
D. L. Clements\altaffilmark{10},
A. Cooray\altaffilmark{11},
H. Dannerbauer\altaffilmark{12},
S. Eales\altaffilmark{3},
E. Ibar\altaffilmark{13},
H. Messias\altaffilmark{14},
M.~J.~Micha{\l}owski\altaffilmark{1}
I. P\'erez-Fournon\altaffilmark{15,16}
E. van Kampen\altaffilmark{2},
}

\affil{$^1$Institute for Astronomy, University of Edinburgh, Royal Observatory, Blackford Hill, Edinburgh EH9 3HJ UK}
\affil{$^2$European Southern Observatory, Karl-Schwarzschild-Str. 2, 85748 Garching, Germany}
\affil{$^3$School of Physics and Astronomy, Cardiff University, The Parade, Cardiff CF24 3AA, UK}
\affil{$^4$Centre for Extragalactic Astronomy, Department of Physics, Durham University, South Road, Durham DH1 3LE UK}
\affil{$^5$Institute for Computational Cosmology, Department of Physics, Durham University, South Road, Durham DH1 3LE, UK}
\affil{$^6$2Department of Astronomy, Cornell University, Ithaca, NY 14853, USA}
\affil{$^7$Department of Physical Sciences, The Open University, Milton Keynes, MK7 6AA, UK}
\affil{$^8$Leiden Observatory, Leiden University, P.O. Box 9513, NL-2300 RA Leiden, The Netherlands}
\affil{$^9$H.H. Wills Physics Laboratory, University of Bristol, Tyndall Avenue, Bristol BS8 1TL, UK}
\affil{$^{10}$Physics Department, Blackett Lab, Imperial College, Prince Consort Road, London SW7 2AZ, UK}
\affil{$^{11}$Department of Physics and Astronomy, University of California, Irvine, CA 92697}
\affil{$^{12}$Universit{\"a}t Wien, Institut f{\"u}r Astrophysik, T{\"u}rkenschanzstrasse 17, 1180 Wien, Austria}
\affil{$^{13}$Instituto de F\'isica y Astronom\'ia, Universidad de Valpara\'iso, Avda. Gran Breta\~na 1111, Valpara\'iso, Chile}
\affil{$^{14}$Centro de Astronomia e Astrof\'isica da Universidade de Lisboa, Observat\'orio Astronómico de Lisboa, Tapada da Ajuda, 1349-018, Lisbon, Portugal}
\affil{$^{15}$Instituto de Astrof\'isica de Canarias, E-38205 La Laguna, Tenerife, Spain}
\affil{$^{16}$ Universidad de La Laguna, Dpto. Astrof\'isica, E-38206 La Laguna, Tenerife, Spain}
\email{ivanoteogomez@gmail.com}


  
\begin{abstract}

Exploiting the sensitivity and spatial resolution of the Atacama Large
Millimeter/submillimeter Array (ALMA), we have studied the morphology
and the physical scale of the interstellar medium - both gas and dust
- in SGP38326, an unlensed pair of interacting starbursts at $z
= 4.425$. SGP38326 is the most luminous star bursting system known at $z > 4$ with an IR-derived ${\rm SFR \sim 4300 \,}
M_\odot \, {\rm yr}^{-1}$. SGP38326 also contains a molecular gas
reservoir among the most massive ever found in the early Universe, and
it is the likely progenitor of a massive, red-and-dead elliptical
galaxy at $z \sim 3$. Probing scales of $\sim 0.1''$ or $\sim 800 \,
{\rm pc}$ we find that the smooth distribution of the continuum
emission from cool dust grains contrasts with the more irregular
morphology of the gas, as traced by the [CII] fine structure
emission. The gas is also extended over larger physical scales than
the dust. The velocity information provided by the resolved [CII]
emission reveals that the dynamics of the two components of SGP38326
are compatible with disk-like, ordered rotation, but also
  reveals an ISM which is turbulent and
  unstable. Our observations support a scenario
  where at least a subset of the most distant extreme starbursts are
  highly dissipative mergers of gas-rich galaxies. 
   
%


\end{abstract}



  
%

\section{Introduction}\label{intro}

Surveys in the far-infrared (FIR) and (sub)-mm have revolutionized our
understanding of the formation and evolution of galaxies, revealing a
population of dusty starbursts at $z > 1$ that are forming stars at
tremendous rates
\citep{Blain2002PhR...369..111B,Chapman2005ApJ...622..772C,Weiss2009ApJ...707.1201W,Geach2013MNRAS.432...53G,Casey2013MNRAS.436.1919C}. At
$z > 4$ these starbursts can be linked to the formation of the
so-called red-sequence -- passively evolving, early type galaxies
which dominate the cores of clusters out to at least $z \sim 1.5$
\citep[e.g.][]{Stanford2006ApJ...646L..13S,Rosati2009A&A...508..583R,Mei2009ApJ...690...42M,Strazzullo2010A&A...524A..17S,Tozzi2015ApJ...799...93T}. While
the bulk of star formation in the {\it general} population of galaxies
occurred in the period since $z \sim 2$
\citep{Sobral2013MNRAS.428.1128S}, red-sequence galaxies at $z > 1.5$
formed the bulk of their stellar population at much earlier times and
over a brief time interval
\citep{Thomas2005ApJ...621..673T,Thomas2010MNRAS.404.1775T}. Therefore,
analysing strong starbursts at $z > 4$ is a way to study the likely
progenitors of the most massive elliptical galaxies at $z > 1.5$ and,
consequently, the birth of the galaxy red sequence, which might have
appeared as early as $z > 2$
\citep{Kodama2007MNRAS.377.1717K,Zirm2008ApJ...680..224Z,Kriek2008ApJ...682..896K,Gabor2012MNRAS.427.1816G,Hartley2013MNRAS.431.3045H}.

{\it Herschel} extragalactic imaging surveys such as {\em Herschel}-ATLAS
\citep{Eales2010PASP..122..499E} and HerMES
\citep{Oliver2010A&A...518L..21O} have covered about $1000 \, {\rm
  deg}^2$ to the SPIRE confusion limit. As a consequence, the number
of known dusty starbursts has increased from several hundred to several
hundred thousand. Most of these are at $z < 3$, but there is also a
population of sources at $z > 4$. Dusty starbursts at $z > 4$ can be
found by looking for galaxies whose far-IR (FIR) spectral energy
distributions (SEDs) rise from 250$\mu$m to 500$\mu$m, so their
thermal dust emission peak is redshifted close to, or beyond
500$\mu$m. These are called 500$\mu$m risers. Using {\it H}-ATLAS
data, we have built a sample of dusty starbursts at $z > 4$ whose SPIRE
flux densities satisfy $S_{\rm 500\mu m} / S_{\rm 250 \mu m} > 2$ and
$S_{\rm 500\mu m} / S_{\rm 350 \mu m} > 1$, and we refer to these as ultrared starbursts (Ivison et al. in prep.). The resulting population
was inspected by eye in each SPIRE band to exclude blended sources,
checked for contamination by radio-loud AGN and correlated with deep
WHT/VISTA/Gemini optical/NIR imaging to reject any lenses that might
have crept in despite our low median $S_{\rm 500\mu m}$ of $\sim 50 \,
{\rm mJy}$. Over 150 of these 500$\mu$m risers were followed up with
SCUBA-2 \citep{Holland2013MNRAS.430.2513H} and LABOCA
\citep{Siringo2009A&A...497..945S} to improve FIR photometric
redshifts and select only those whose colors are consistent with $z >
4$ by imposing $S_{\rm 870 \mu m} / S_{\rm 500 \mu m} > 0.4$. 

In this paper we present ALMA high-spatial resolution ($\sim 0.1''$)
observations of dust and gas (traced by [CII] emission) in HATLAS J000307$-$330250, also known as SGP38326,
one of our ultra-red starbursts confirmed to be at $z =
4.425$ via multi-CO line detection. In \S \ref{sec_desctiption_ALMA_data} we give a description of the ALMA data used in this work and how they were reduced and analyzed. In \S \ref{descriptopn_source_this_work} we present and discuss the main results of this work, regarding the properties of SGP38326: detected emission lines, the implied molecular gas mass, morphology of dust and atomic gas, etc. Finally, \S \ref{concluuuuu} summarizes the main conclusions of this work. The total IR luminosities ($L_{\rm
  IR}$) reported in this work refer to the integrated luminosities
between rest-frame 8 and 1000$\mu$m. Throughout this paper, the
reported SFRs are derived from the total IR luminosities ($L_{\rm IR,
  8-1000\mu m}$) assuming a Salpeter IMF and the
\cite{Kennicutt1998ARA&A..36..189K} calibration. Unless explicitly specified, the areas are calculated by carrying out 2D elliptical gaussian fits and using $A = \pi \times R_1 \times R_2$, where $R_1$ and $R_2$ are the semi axes of the best-fitting elliptical gaussians, $R_1 = {\rm FWHM}_1/2$ and $R_2 = {\rm FWHM}_2/2$. All values taken from the literature are re-scaled to be consistent. We
assume a flat universe with $(\Omega_m, \Omega_\Lambda, h_0)=(0.3,
0.7, 0.7)$, and all magnitudes are listed in the AB system
\citep{Oke1983}. For this cosmology, the sky scale is $\sim 6.6\,{\rm
  kpc}/''$ at $z = 4.425$, the redshift of SGP38326. 

\section{ALMA data}\label{sec_desctiption_ALMA_data}

This work makes use of ALMA data from two different projects. Project
2013.1.00449.S was aimed at deriving redshifts via spectral scans in
the ALMA 3mm window (PI. A. Conley) of a sample of ultra-red
starbursts at $z > 4$ of which SGP38326 was a part. Project
2013.1.00001.S was a high-spatial resolution study of the dust
continuum morphology at 870$\mu$m of a subsample of our ultra-red
starbursts (PI. R.J. Ivison) of which SGP38326 was also a part. The ALMA data corresponding to the two projects
were calibrated by using the ALMA pipeline and by executing different
ALMA calibration scripts depending on the date when the data were
released. The calibrated visibilities for the calibrators and science
target were visually inspected and only very minor further flagging
was required. The calibrated visibilities of the science target were
then imaged in CASA by using a {\sc natural} weighting scheme to
maximize sensitivity. The different tunings of the ALMA 3mm spectral
scan were carried out at different dates with the array in different
configurations. Therefore, the beam size is different for each
tuning. In this work we only use the ALMA 3mm data corresponding to
the tunings where the $^{12}$CO(4--3) and $^{12}$CO(5--4) lines were
detected. The beam sizes were $1.5'' \times 1.2''$ and $1.1'' \times
1.0''$ for the $^{12}$CO(4--3) and $^{12}$CO(5--4) transitions,
respectively. The r.m.s. of the $^{12}$CO(4--3) and $^{12}$CO(5--4)
observations are $0.54 \, {\rm mJy \, beam^{-1}}$ and $0.64 \, {\rm
  mJy \, beam^{-1}}$, respectively, in $50 \, {\rm km/s}$
channels. The beam size of the 870$\mu$m observations is $0.16''
\times 0.12''$, which corresponds to a physical scale of about $1.0 \,
       {\rm kpc} \times 790 \, {\rm pc}$. The r.m.s. of the 870$\mu$m
       continuum map is $0.11 \, {\rm mJy \, beam^{-1}}$, while the
       r.m.s. in the spectral window where [CII] is detected is $0.66
       \, {\rm mJy \, beam^{-1}}$ in $100 \, {\rm km/s}$ channels.

\section{SGP38326: a pair of interacting starbursts at $z = 4.425$}\label{descriptopn_source_this_work}

\begin{figure}[!t]
\centering
\includegraphics[width=0.48\textwidth]{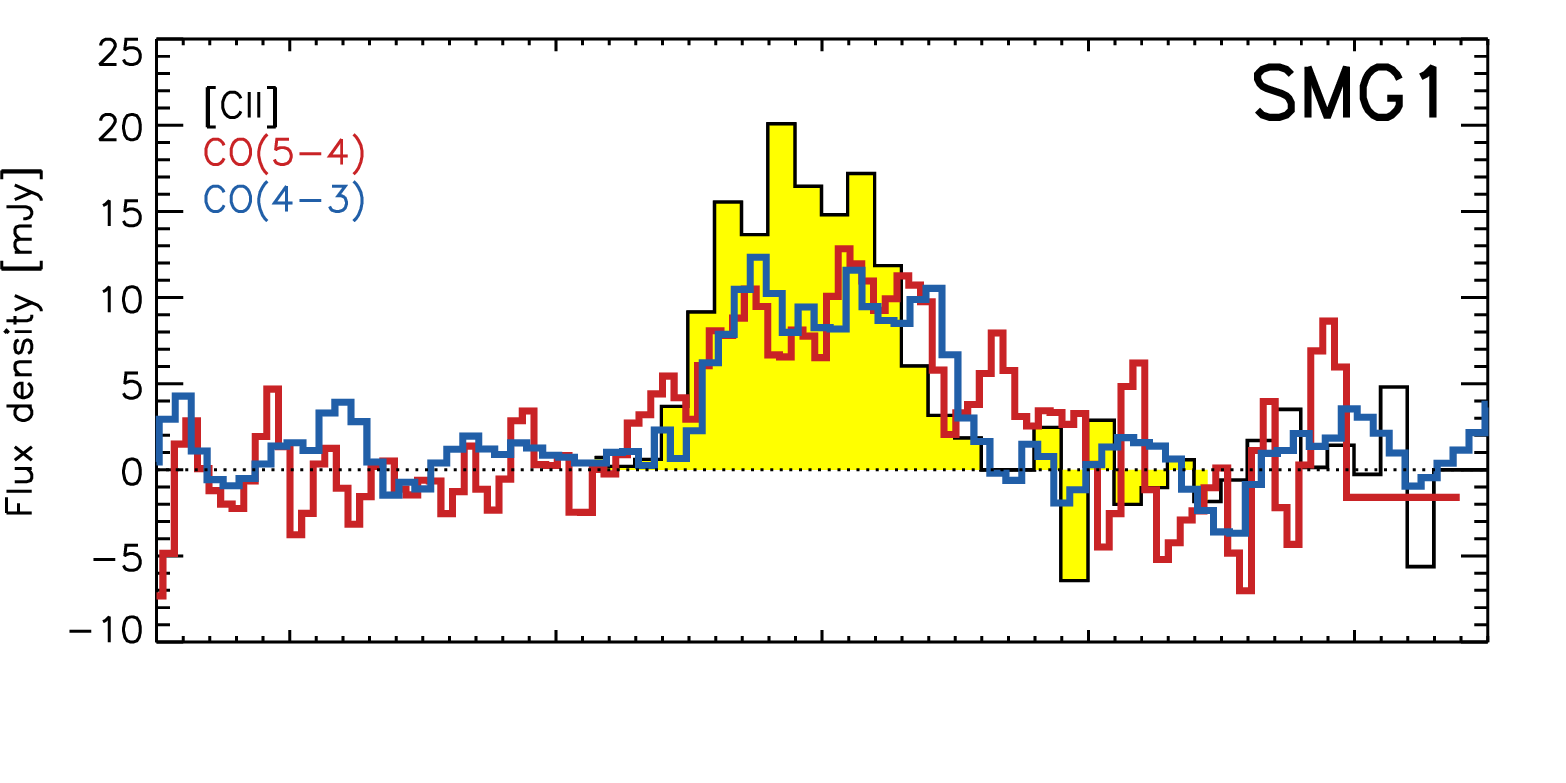}\\
\vspace{-9mm} 
\includegraphics[width=0.48\textwidth]{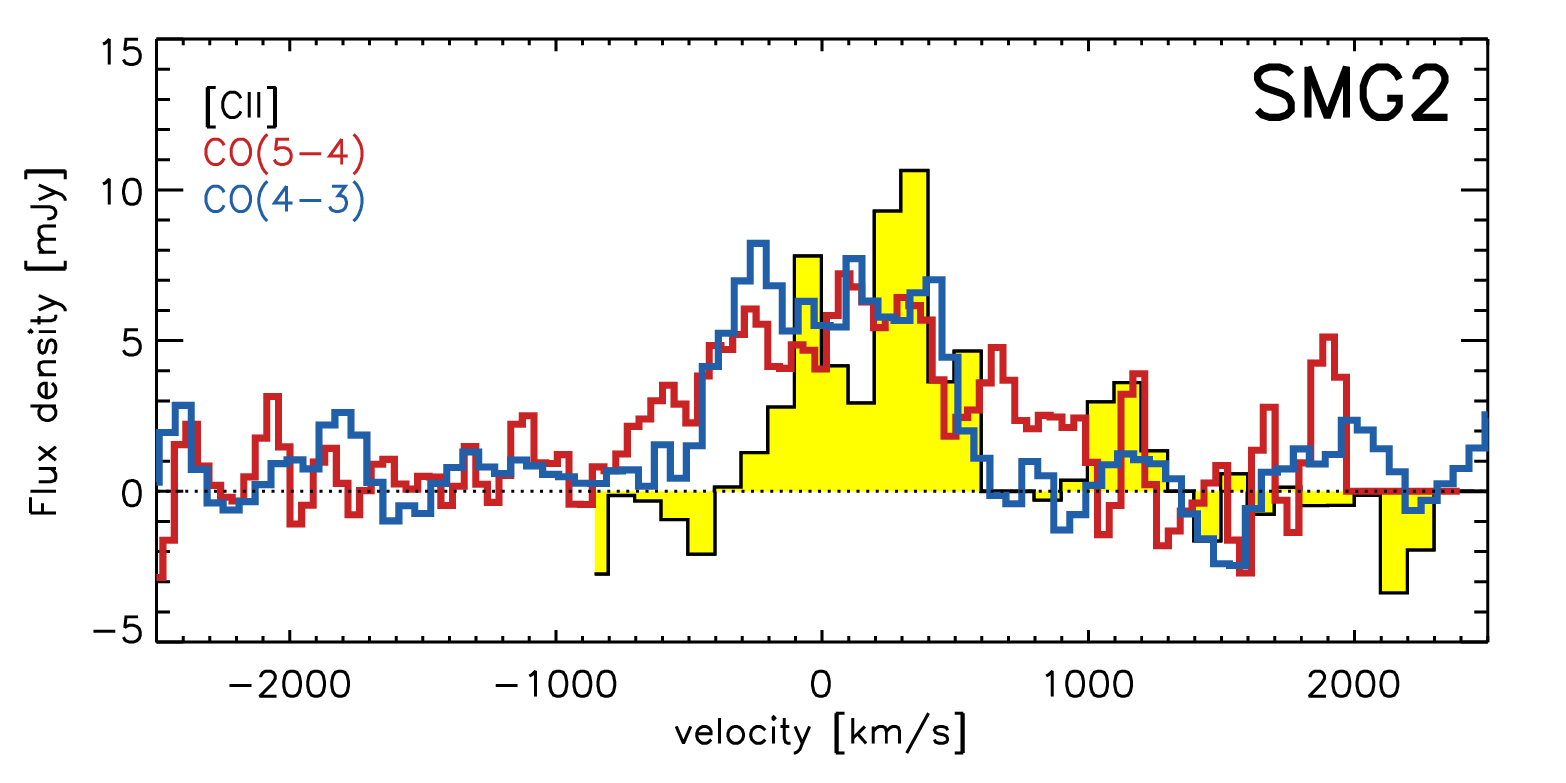}\\
\includegraphics[width=0.48\textwidth]{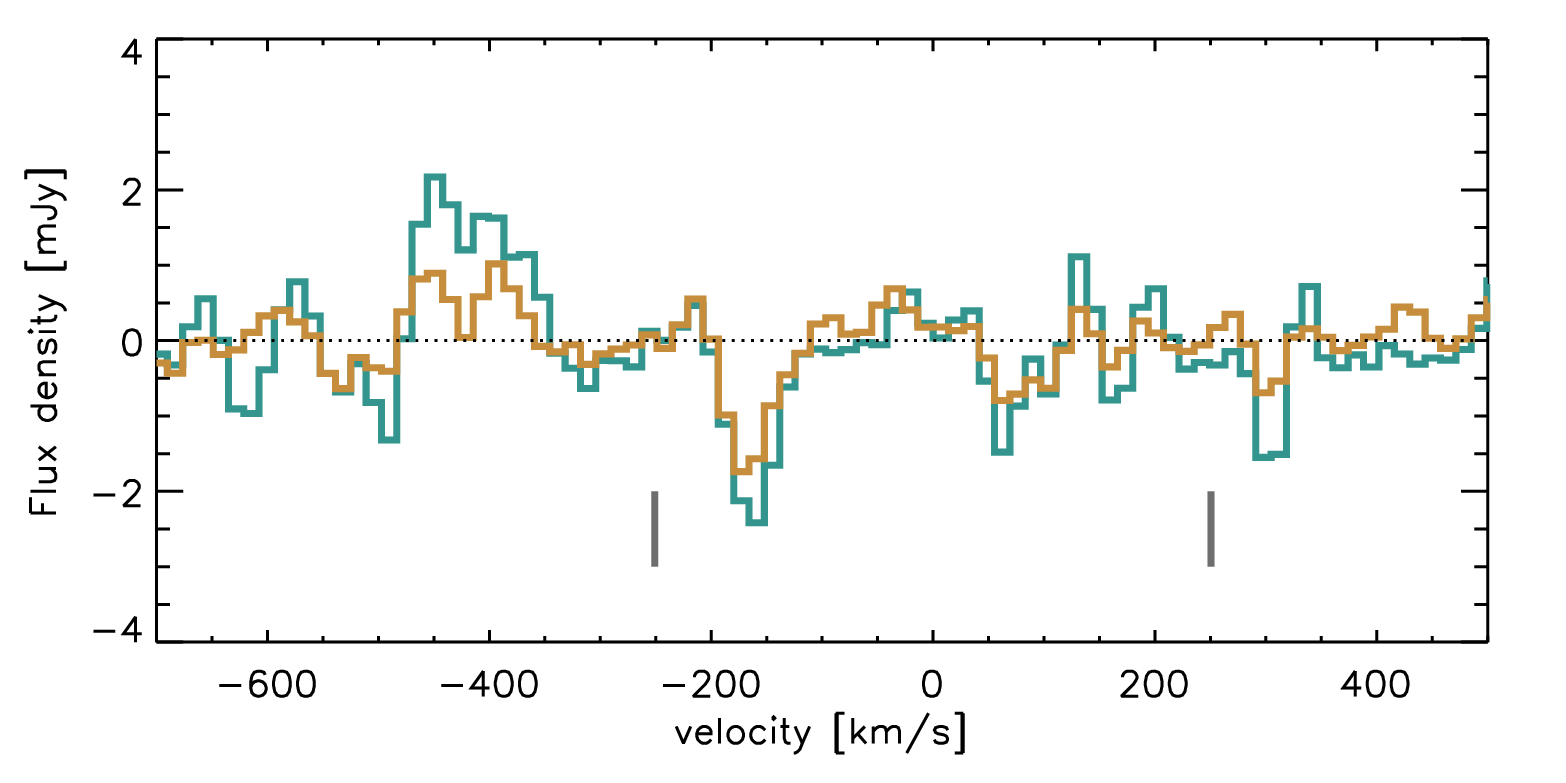}
\caption{\emph{Upper and middle}: Unambiguous confirmation that
  SGP38326 lies at $z = 4.425$ via detection of $^{12}$CO(5--4) and
  $^{12}$CO(4--3) in an ALMA 3mm spectral scan. The CO lines shown correspond to the emission of SMG1 and SMG2. The [CII] emission of
  the two interacting components of SGP38326 (see text for more
  details) are also shown. The CO spectra have been re-scaled
  (multiply by 3 for $^{12}$CO(5--4) and 6 for $^{12}$CO(4--3)) to
  match the flux scale of the [CII] line. \emph{Bottom}: Faint (or
  absent) OH 163$\mu$m emission in SMG1. The green curve represents
  the spectrum in the center of SMG1 (extracted from an aperture equal
  to one synthesized beam), where dust emission is at its maximum. The
  brown spectrum has been extracted using an aperture enclosing the
  whole area where [CII] emission is detected. Small vertical grey
  lines represent where the two components of the OH 163$\mu$m would
  be located assuming the redshift derived from the detected CO and
  [CII] lines. There is a narrow ($ 81 \pm 16 \, {\rm km \, s^{-1}}$,
  FWHM) emission line at $\sim -400\,{\rm km \, s^{-1}}$ at the center
  of SMG1 (where the dust continuum emission is at its maximum) that
  we tentatively associate to OH 163$\mu$m emission (see details in \S
  \ref{lack_OH_emission}). No OH 163 $\mu$m emission or absorption is
  detected in SMG2 either (spectrum not shown for the sake of
  representation clarity). In all panels, the
  spectra shown were extracted after continuum subtraction.  }
\label{spec_CO_CII_lines}
\end{figure}

\begin{figure}[!t]
\centering
\includegraphics[width=0.48\textwidth]{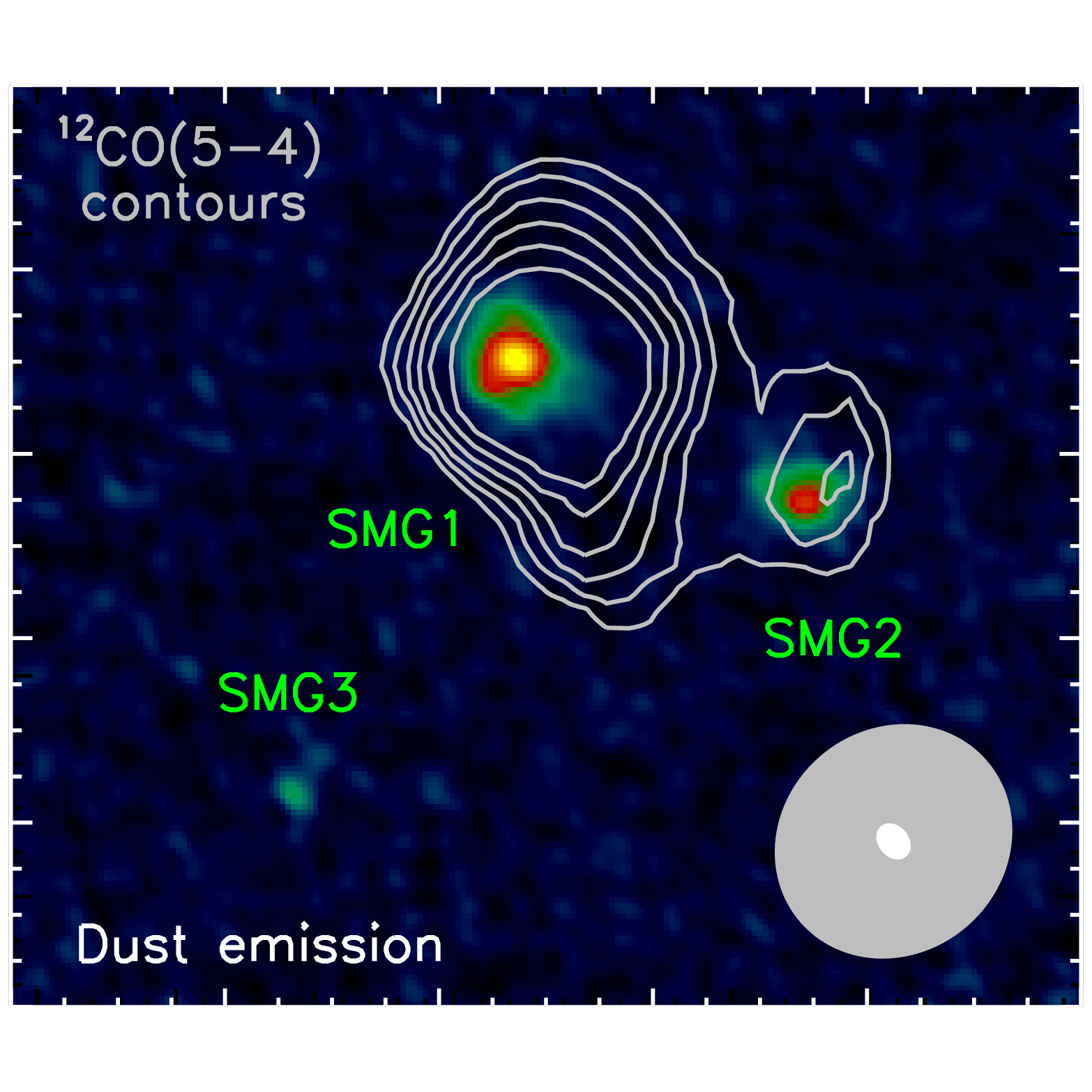}\\
\vspace{-7mm}
\includegraphics[width=0.48\textwidth]{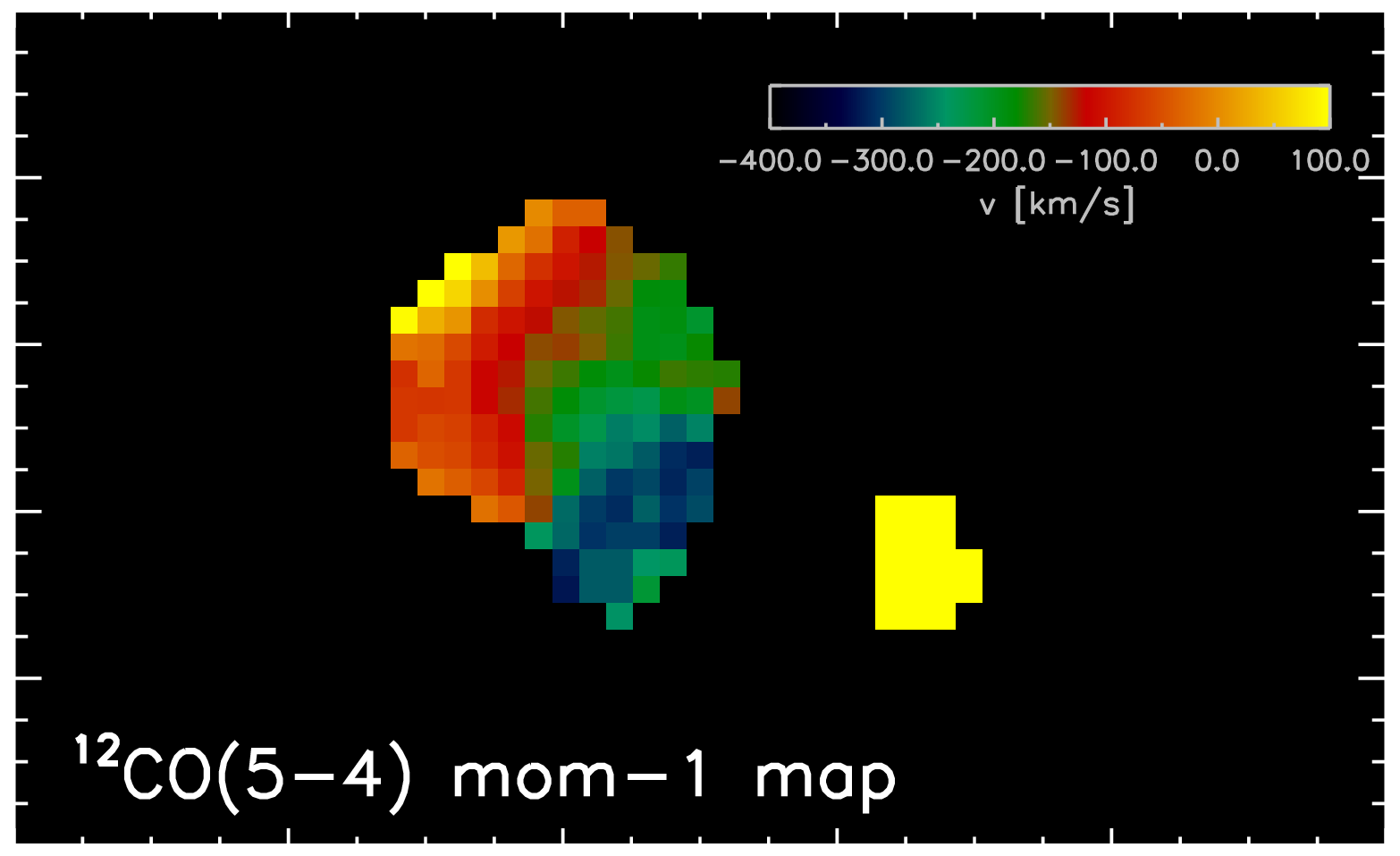}
\caption{\emph{Upper}: ALMA 870$\mu$m continuum map of SGP38326. The
  three detected SMGs are indicated. The grey contours represent the
  integrated $^{12}$CO(5--4) emission. This line is clearly detected
  in both SMG1 and SMG2, but there is no $^{12}$CO(5--4) detection in
  SMG3. The beam sizes of the $^{12}$CO(5--4) (grey ellipse,
  $1.1'' \times 1.0''$) and dust continuum (white ellipse, $0.16''
  \times 0.12''$) observations are shown, and clearly highlight the
  impressive increase in spatial resolution. \emph{Bottom}: Velocity
  map of SMG1 derived from the $^{12}$CO(5--4) emission using moment
  masking \citep{Dame2011arXiv1101.1499D}. It can be seen that,
  despite the lack of spatial resolution, the $^{12}$CO(5--4)
  observations already indicate that SMG1 presents a disk-like
  rotation. In both panels, north is up and east is left.  }
\label{dust_CO_morph_3comp}
\end{figure}

\begin{figure}[!t]
\centering
\includegraphics[width=0.48\textwidth]{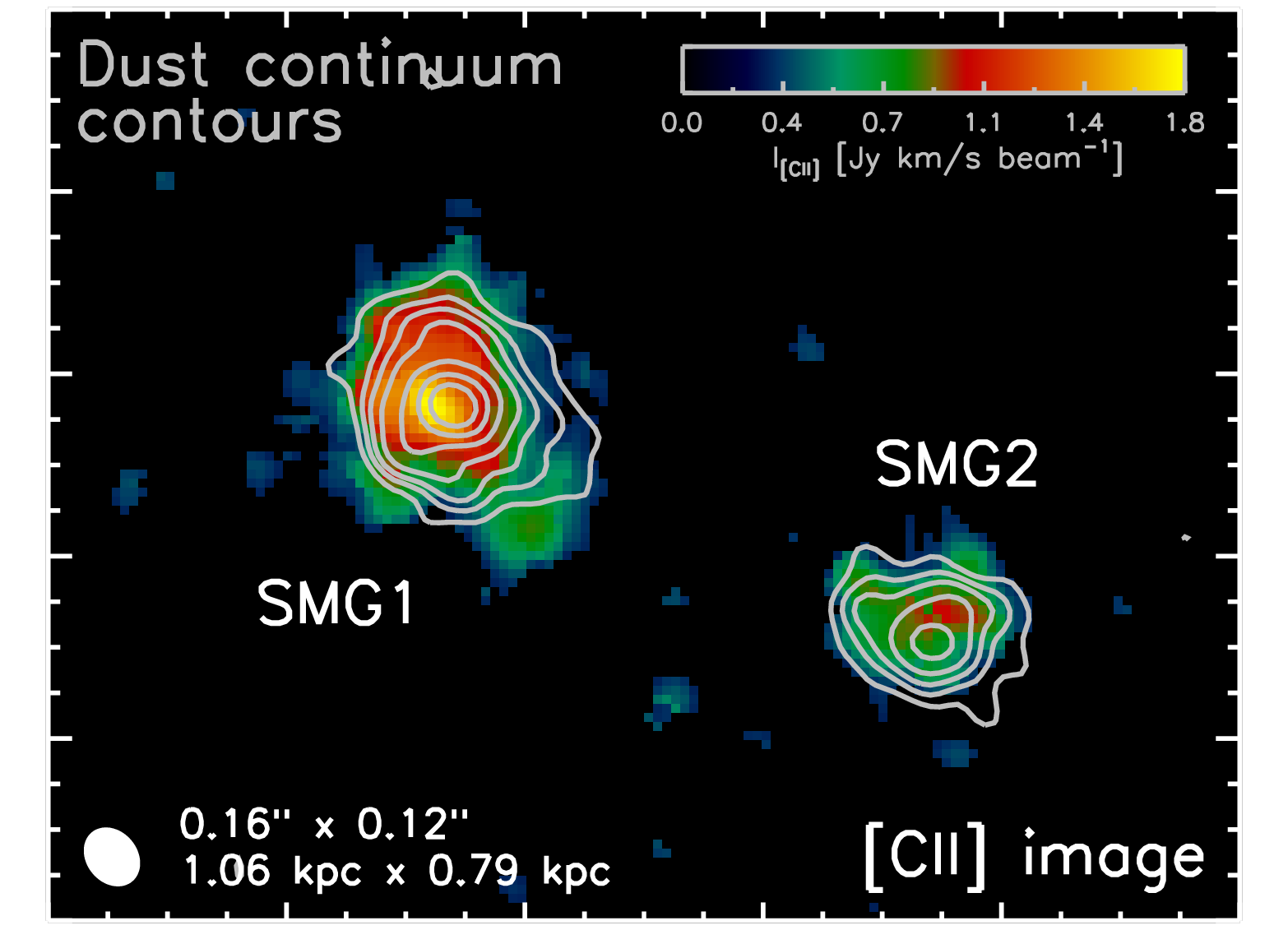} \\
\includegraphics[width=0.22\textwidth]{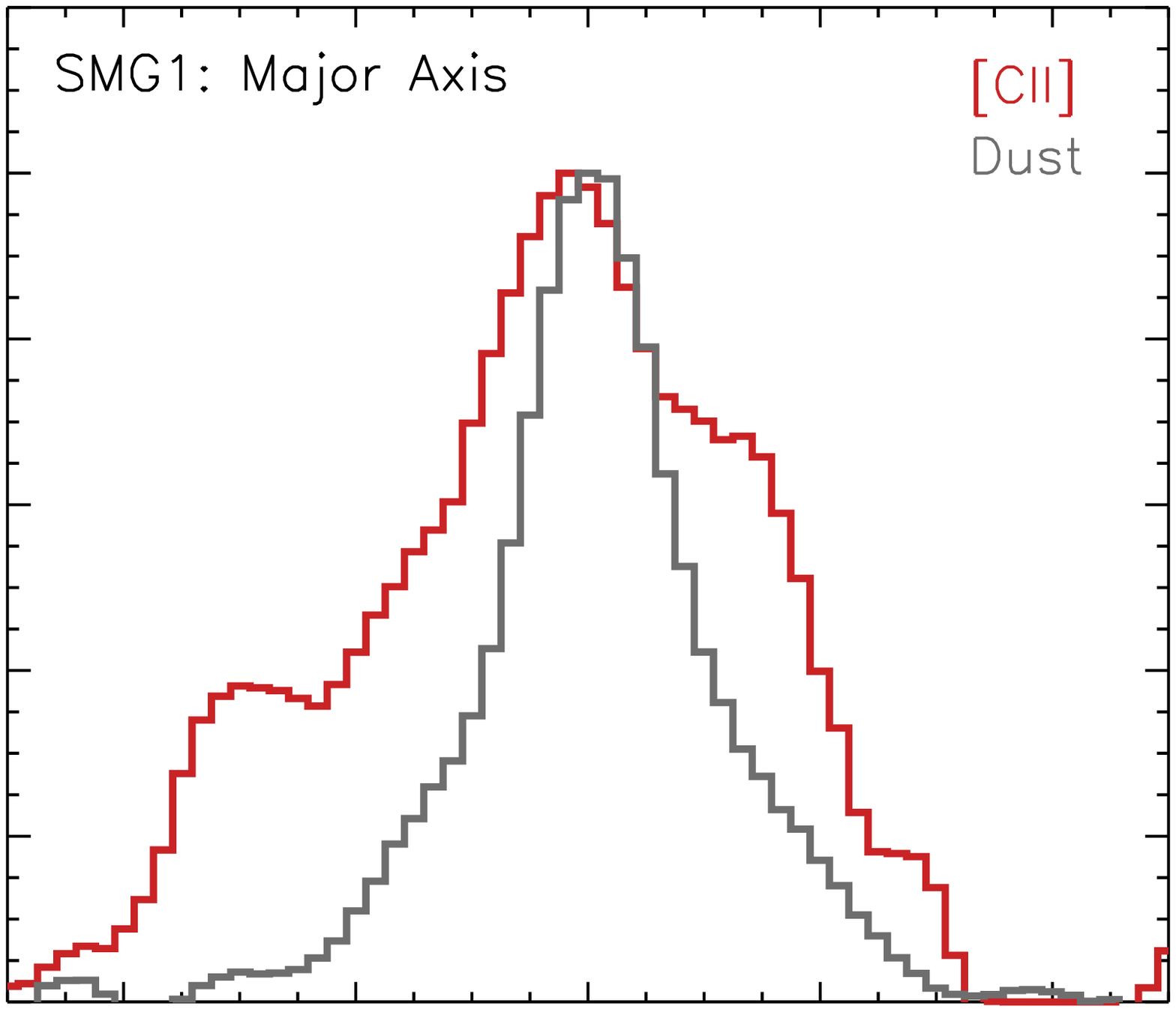} 
\hspace{-2mm}
\includegraphics[width=0.22\textwidth]{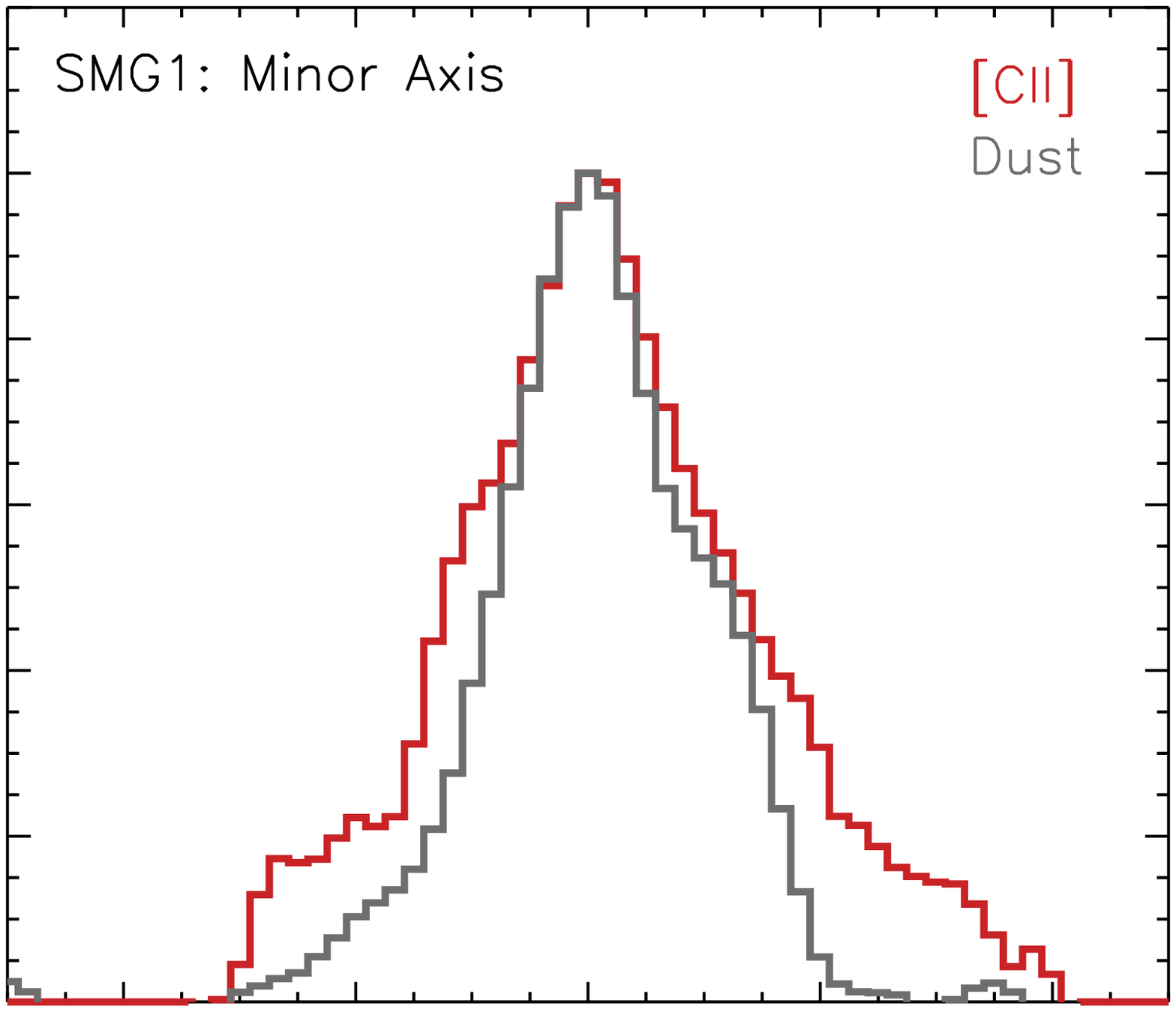}\\
\vspace{-6mm}
\includegraphics[width=0.22\textwidth]{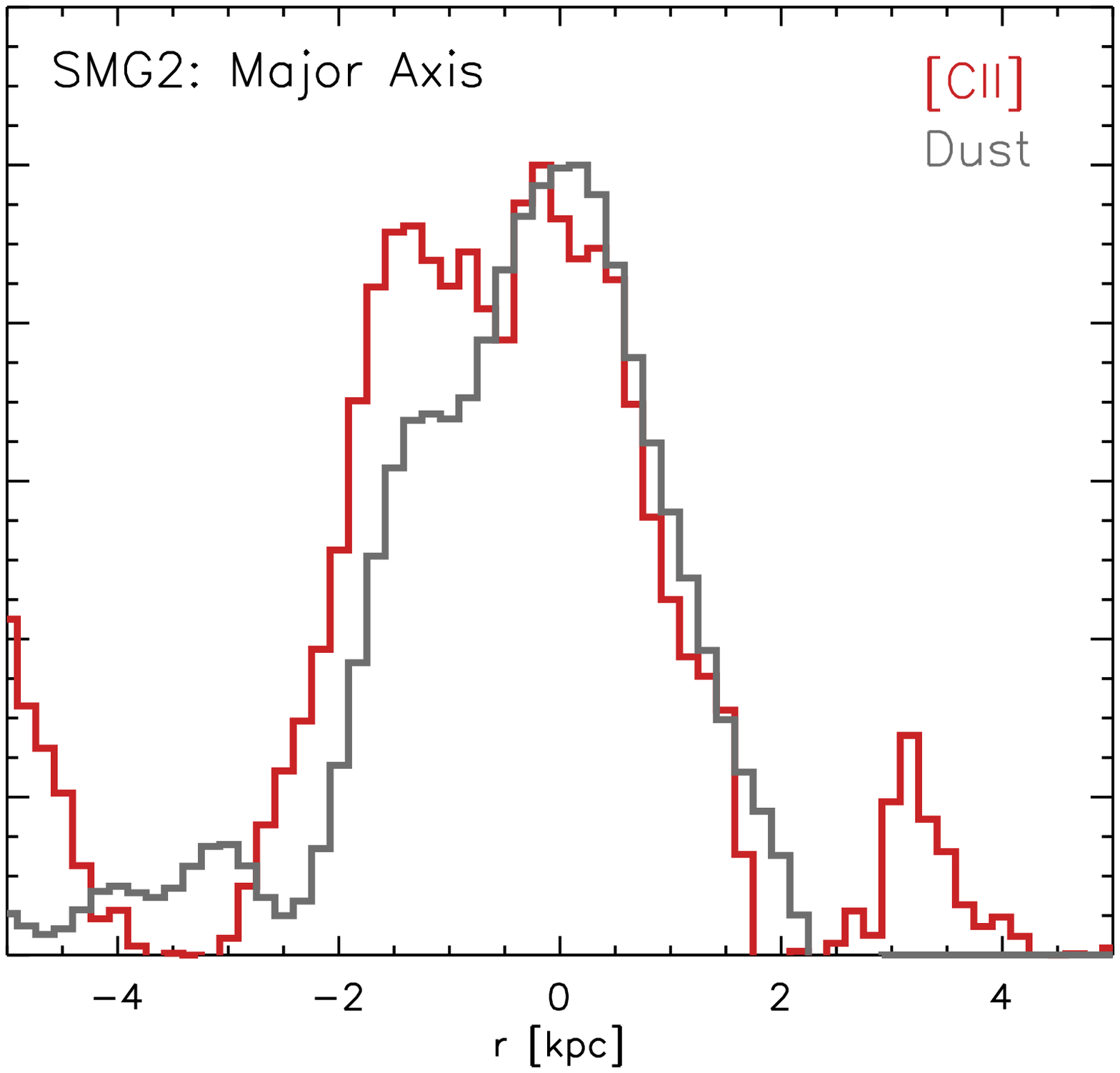} 
\hspace{-2mm}
\includegraphics[width=0.22\textwidth]{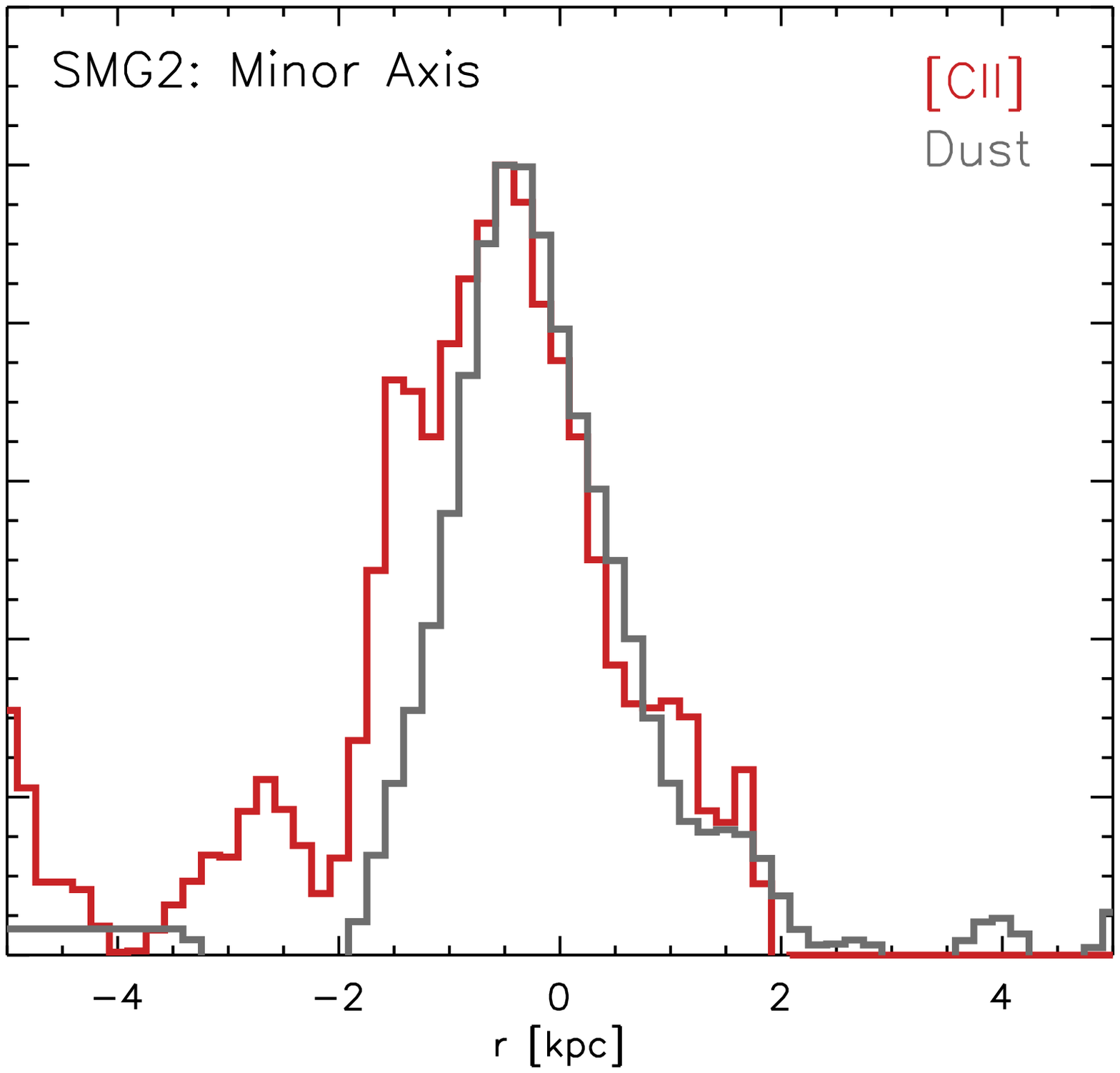}\\
\caption{Morphology and spatial extent of the gas and dust in our two distant interacting starbursts. In the top panel, the background image is the integrated [CII] emission created by using moment masking \citep{Dame2011arXiv1101.1499D}, while grey contours represent dust emission at rest-frame 160$\mu$m. Contours are plotted from 3$\sigma$ to 7$\sigma$ in steps of 2$\sigma$ and from 10$\sigma$ to 25$\sigma$ in steps of 5$\sigma$, where $\sigma = 0.11 \, {\rm mJy \, beam^{-1}}$. The two interacting galaxies are separated by a physical distance of $\sim 10 \, {\rm kpc}$. The beam size is also included. It can be seen that the dust continuum is smooth at the resolution of these data. However, the gas reservoir (as traced by [CII] emission) has an irregular morphology and is extended over larger physical scales. In this panel, north is up and east is left. The bottom panels shows the radial profiles (normalized units) of dust and gas along the [CII] major and minor axes of each starburst. It can be seen that the profiles are significantly different, with [CII] emission being more extended than dust, especially in SMG1.
              }
\label{radial_profiles_dust_CII_figure}
\end{figure}

SGP38326 was selected in the Southern field of the {\it Herschel}-ATLAS
\citep[][Valiante et al. in prep]{Eales2010PASP..122..499E} as a 500$\mu$m riser (see \S
\ref{intro}). The SPIRE flux densities of SGP38326 are $S_{\rm 250 \mu
  m} = 28.6 \pm 5.2 \, {\rm mJy}$, $S_{\rm 350 \mu m} = 28.6 \pm 5.8
\, {\rm mJy}$, $S_{\rm 500 \mu m} = 46.2 \pm 6.8 \, {\rm mJy}$. SGP38326 was
subsequently observed with SCUBA-2 at 870 $\mu$m ($S_{\rm 870 \mu m} = 32.5 \pm 4.1 \, {\rm
  mJy}$) and, by fitting a set of FIR templates to
its SPIRE and SCUBA-2 flux densities, we determined a best-fit photometric redshift
$z_{\rm phot} \sim 4.5$.

The ALMA spectral scan in the 3mm window confirmed its redshift to be
$z = 4.425 \pm 0.001$ via detection of the $^{12}$CO(4--3) and
$^{12}$CO(5--4) emission lines (Fig.~\ref{spec_CO_CII_lines}). The
total IR luminosity of SGP38326, obtained from an SED fit to the {\it Herschel} and SCUBA-2 flux densities, is
$L_{\rm IR} / L_\odot \sim 2.5 \times 10^{13}$, implying a staggering
IR-derived ${\rm SFR} \sim 4300 \, M_\odot \, {\rm yr}^{-1}$, among
the strongest starbursts so far found in the early Universe (see
Table \ref{table_comparison_sources}). The CO emission lines are
extremely wide, ${\rm FWHM_{^{12}CO(4-3)} = 1081 \pm 91 \, {\rm km \,
    s^{-1}}}$ and ${\rm FWHM_{^{12}CO(5-4)} = 1278 \pm 84 \, {\rm km
    \, s^{-1}}}$, from gaussian fits. This is compatible with SGP38326
being a system of at least two starbursts in the process of
interaction, as already seen in extreme starbursts at lower redshifts
with similarly wide CO lines
\citep{Fu2013Natur.498..338F,Ivison2013ApJ...772..137I}. In fact, the
$^{12}$CO(5--4) observations resolved SGP38326 into two interacting
components, SMG1 and SMG2 (see Figure \ref{dust_CO_morph_3comp}). The
$^{12}$CO(5--4) emission in SMG1 presents an elongated extention
towards the south. This can be interpreted as a faint and relatively
dust poor CO emitter interacting with SMG1 and SMG2, a CO bridge
connecting the two interacting components (see also later in the
text), or an outflow. The depth and resolution of the data are
insufficient to fully discriminate between the possible
scenarios. As discussed later, the [CII] velocity map shows no
evidence of outflowing atomic gas and OH 163$\mu$m emission (a tracer of molecular outflows) is not
clearly detected in SMG1 despite the detection of the continuum at the
line frequency. The existence of a CO bridge is supported by the
detection of similar structures in other systems of interacting
starbursts at lower redshifts
\citep{Zhu2007AJ....134..118Z,Fu2013Natur.498..338F,Appleton2015ApJ...812..118A}
and QSOs \citep{Carilli2013ApJ...763..120C}. The velocity map of SMG1
obtained from the $^{12}$CO(5--4) emission is compatible with an
ordered rotation (see Figure \ref{dust_CO_morph_3comp}). As
shown in \S \ref{dynamics_from_CII_line}, this velocity structure is
in agreement with that obtained from the [CII] emission at $\sim 10$x
times better spatial resolution (see \S \ref{dynamics_from_CII_line}).

\begin{table*}[!t]
\begin{center}
\caption{Properties of SGP38326 compared to other IR-bright starbursts in the distant Universe.}
\label{table_comparison_sources}
\begin{tabular}{l c c c c c c c c}
\hline
Source	&	redshift	& $S_{\rm 870 \mu m} \, [{\rm mJy}]$		& SFR [$M_\odot \, {\rm yr}^{-1}$] & $M_{\rm H_2} \, [M_\odot]$\tablenotemark{a} & $\Sigma_{\rm SFR} \, [M_\odot \, {\rm yr}^{-1} \, {\rm kpc}^{-2}]$ & Lensed?\tablenotemark{b} & Reference \\
\hline
SGP38326	&	4.425	& $32.5 \pm 4.1$ &	$\sim 4300$\tablenotemark{c}	&	$\sim 2.7 \times 10^{11}$	&  $\sim 840$ / $\sim 570$\tablenotemark{d} & No	& This work	\\
GN20		&	4.055	& $20.3 \pm 2.1$ &	$\sim 3000$	&	$\sim 1.3 \times 10^{11}$	&	$\sim 310$	 & No & 	\cite{Hodge2012ApJ...760...11H}	\\
HDF\,850.1	&	5.183	& $7.0 \pm 0.4$&	$\sim 850$		&	$\sim 3.5 \times 10^{10}$	& 	$\sim 50$	& Weakly & 	\cite{Walter2012Natur.486..233W}					\\
AzTEC-3		&	5.299	& $6.20 \pm 0.25$\tablenotemark{e}&	$\sim 1980$	&	$\sim 5.3 \times 10^{10}$	& $\sim 840$		& No & \cite{Riechers2014ApJ...796...84R}	\\
HFLS3		&	6.337	& $14.3 \pm 1.1$&	$\sim 2100$	&	$\sim 3.6 \times 10^{10}$ 	& $\sim 600$	& Weakly		& \cite{Riechers2013Natur.496..329R}	\\

\hline
\end{tabular}
\tablenotetext{1}{Molecular gas masses reported assume an $\alpha_{\rm CO} = M_{\rm gas} / L^{'}_{\rm CO} = 0.8 \, {\rm K \, km \, s^{-1} \, pc^{2}}$, typical for high-redshift starbursts and local ULIRGs \citep{Downes1998ApJ...507..615D}. Values taken from the literature have been scaled accordingly.}
\tablenotetext{2}{For lensed galaxies, de-magnified values are listed.}
\tablenotetext{3}{Total SFR in SGP38326: ${\rm SFR = SFR_{SMG1} + SFR_{SMG2}}$.}
\tablenotetext{4}{Values for SMG1 and SMG2 are given.}
\tablenotetext{5}{Measured at 1mm.}
\end{center}
\end{table*}

SGP38326 is resolved into three 870$\mu$m components (see Figure
\ref{dust_CO_morph_3comp}), as are many bright SMGs
\citep{Karim2013MNRAS.432....2K,Hodge2013ApJ...768...91H,Simpson2015ApJ...799...81S,Simpson2015ApJ...807..128S,Bussmann2015ApJ...812...43B}. SMG1
and SMG2 (Fig.~\ref{radial_profiles_dust_CII_figure}), the
brightest components (see flux densities in Table
\ref{table_values_individual_sources}), are both at the same redshift,
$z = 4.425$. There is no [CII] or CO emission detected in SMG3
($S_{\rm 870 \mu m}^{\rm SMG3} = 1.7 \pm 0.4 \, {\rm mJy}$), located
$2.2''$ away from SMG1. This might indicate that SMG3 has a different
redshift than SMG1 and SMG2. However, SMG3 is about 14 times fainter
than the SMG1 and SMG2 combined. If we assume that the CO luminosity
scales with the total IR luminosity, the peak flux of the CO lines
would be much lower than the r.m.s. of the observations. The same is true for [CII] assuming again that its luminosity scales with the
total IR luminosity. Therefore, with the current data, is not possible
to clarify whether SMG3 is also interacting with SMG1 and SMG2. From
now on, we focus our study on SMG1 and SMG2, the two clear components
of the merging system. These are separated by $\sim$10 kpc in the
plane of the sky and their properties are summarized in Table
\ref{table_values_individual_sources}.

Due to the high inferred luminosity of SGP38326, we should consider whether its emission has been amplified by gravitational
lensing. SGP38326 has no near-IR counterpart in the VIKING survey \citep{Edge2013Msngr.154...32E}; the
closest VIKING-detected source is $\sim7.5''$ away. The $5 \sigma$
limiting magnitudes of the VIKING survey range from $23.1 \, {\rm
  mag}$ in the $Z$ band to $21.2 \, {\rm mag}$ in the $K_s$ band. Assuming
that any lens galaxy would be an elliptical, the limiting magnitude in $Z$
would imply $I < 23.1 \, {\rm mag}$. Therefore, the lack of VIKING
detection would imply that any potential lensing galaxy would be
significantly fainter than the lenses in the H-ATLAS survey discovered
so far \citep{Bussmann2013ApJ...779...25B}. Additionally, the dust
continuum emission does not show any evidence of lensing, as it does
in some ultra-red starburst at $z > 4$ (Oteo et al. in
prep). Therefore, neither of the two merging components is lensed, in
contrast with other high-$z$ extremely bright galaxies reported in the
literature, such as HFLS3
\citep{Riechers2013Natur.496..329R,Cooray2014ApJ...790...40C},
HDF\,850.1 \citep{Walter2012Natur.486..233W}, HXMM01
\citep{Fu2013Natur.498..338F} or HATLAS\,J084933
\citep{Ivison2013ApJ...772..137I}.

\begin{table}
\caption{Observed properties of our two interacting dusty starbursts}\label{table_values_individual_sources}
\centering
\begin{tabular}{ccc}
\hline
& SMG1 & SMG2 \\
\hline
\hline
$z_{\rm spec}$	& $4.4237 \pm 0.0004$ & $4.4289 \pm 0.0004$ \\
$S_{\rm 870 \mu m}$\tablenotemark{a}& $16.3 \pm 1.1 \, {\rm mJy}$ & $7.3 \pm 0.5 \, {\rm mJy}$ \\
$I_{\rm [CII]} \, [{\rm Jy \, km \, s^{-1}}]$ & $13.9 \pm 1.1$ & $5.3 \pm 0.6$ \\
$\Delta v_{\rm [CII]} \, [{\rm km \, s^{-1}}]$ & $626 \pm 60$ & $585 \pm 95$ \\
$L_{\rm IR}\,[L_\odot]$ & $(1.6 \pm 0.3) \times 10^{13}$ & $(7.9 \pm 0.3) \times 10^{12}$ \\
${\rm SFR \, [M_\odot \, yr^{-1}]}$ & $\sim 2900$ & $\sim 1400$\\
$A_{\rm dust}\,[{\rm kpc^2}]$\tablenotemark{b} & $\rm 2.2\pm0.2 \times 2.0\pm0.2$ & $2.1\pm0.2 \times 1.5\pm0.1$ \\
$A_{\rm [CII]}\,[{\rm kpc^2}]$\tablenotemark{b} & $3.8\pm0.1 \times2.9\pm0.1$ & $2.7\pm0.1 \times 2.1\pm0.1$ \\
$\Sigma_{\rm SFR}\,[{\rm M_\odot \, yr^{-1} \, kpc^{-2}}]$ & $\sim 840$ & $\sim 570$ \\
$M_{\rm H_2} \, [M_\odot]$ & $ \sim 1.9 \times 10^{11}$ & $ \sim 7.6 \times 10^{10}$ \\
$M_{\rm dust}\tablenotemark{c} \, [M_\odot]$ &  $\sim 1.9 \times 10^{9}$ & $\sim 6.9 \times 10^{8}$ \\
$\Sigma_{\rm H_2}\,[{\rm M_\odot \, pc^{-2}}]$ & $\sim 5.6 \times 10^4$ & $\sim 3.1 \times 10^4$ \\
$T_{\rm dust}$\tablenotemark{c} [K] & $\sim 33$ & $\sim 34$\\
$L_{\rm [CII]} \, [L_\odot]$ &  $(8.3 \pm 0.2) \times 10^9$ & $(2.9 \pm 0.2) \times 10^9$\\
\hline
\hline
\end{tabular}
\tablenotetext{1}{Flux densities derived from the ALMA 870$\mu$m observations (\S \ref{sec_desctiption_ALMA_data}) after primary beam attenuation correction. At the high signal to noise of the 870$\mu$m detections, the effect of flux boosting is negligible \citep{Oteo2015arXiv150805099O}.}
\tablenotetext{2}{The reported values of the size of dust and [CII] emission are ${\rm FWHM_{major} \times FWHM_{minor}}$, where ${\rm FWHM_{major}}$ and ${\rm FWHM_{minor}}$ are obtained from a two-dimensional elliptical gaussian profile fit to the observed emission.}
\tablenotetext{3}{Values obtained assuming optically thin models}
\end{table}

\subsection{A massive molecular gas reservoir}\label{section_massive_mol_gas_reser}

The integrated $^{12}$CO(5--4) emission shown in Figure \ref{dust_CO_morph_3comp} can be used to estimate the total mass of the molecular gas reservoir in SGP38326, as well as the molecular gas mass of each interacting component. Low-$J$ CO lines are much better tracers of the total molecular gas mass than the mid-$J$ CO lines (the ones detected in our ALMA 3mm spectral scan), which trace only a relatively dense component. Therefore, we need to assume that the average CO line ratio for SMGs \citep{Carilli2013ARA&A..51..105C}, $L'_{\rm CO(5-4)} / L'_{\rm
  CO(1-0)} = 0.39$, is valid for SGP38326. The molecular gas masses are then
calculated using $\alpha_{\rm CO} = 0.8 M_\odot \, ({\rm K \, km \,
  s^{-1} \, pc^2})$, a value typically assumed for high-$z$ starbursts
and local ULIRGs \citep{Downes1998ApJ...507..615D}. The derived values
are shown in Table \ref{table_values_individual_sources}. The total
molecular gas of SGP38326 ($M_{\rm g, \, total} = M_{\rm g}^{\rm SMG1}
+ M_{\rm g}^{\rm SMG2}$ -- see also Table
\ref{table_comparison_sources}) is higher than that in any other
system so far studied at similar redshifts (see Table
\ref{table_comparison_sources}). We note that the calculation
of the molecular gas mass from a single mid-$J$ transition might be
highly uncertain due to the effect that variation in excitation has on
the CO line ratios. Further observations of low--$J$ CO transitions would thus
be needed to confirm whether the high luminosity of the $^{12}$CO(5--4)
transition is due to the presence of a very massive molecular gas
reservoir rather than a highly excited gas component.


\begin{figure*}[!t]
\centering
\includegraphics[width=0.32\textwidth]{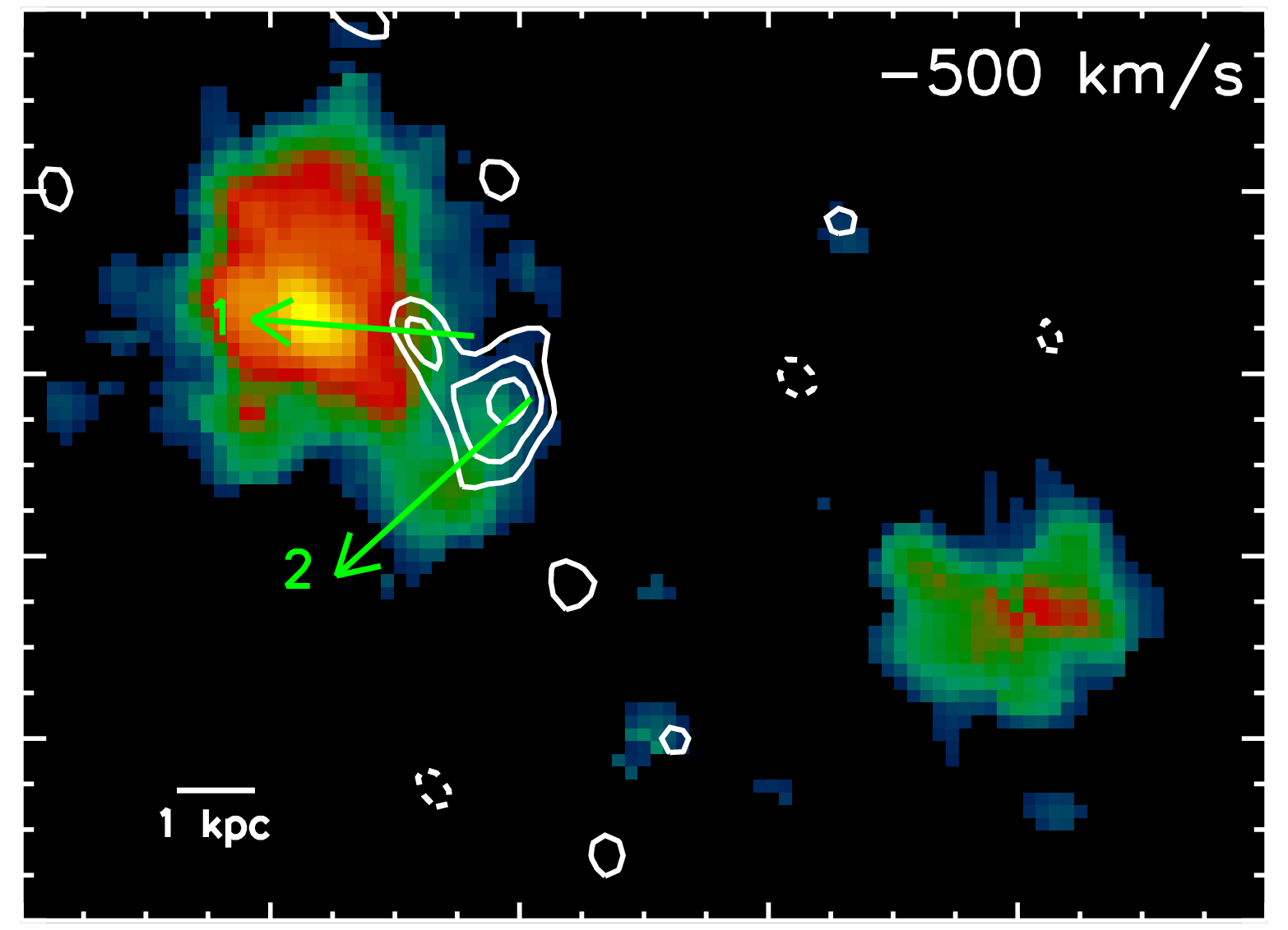} 
\hspace{-2.4mm}
\includegraphics[width=0.32\textwidth]{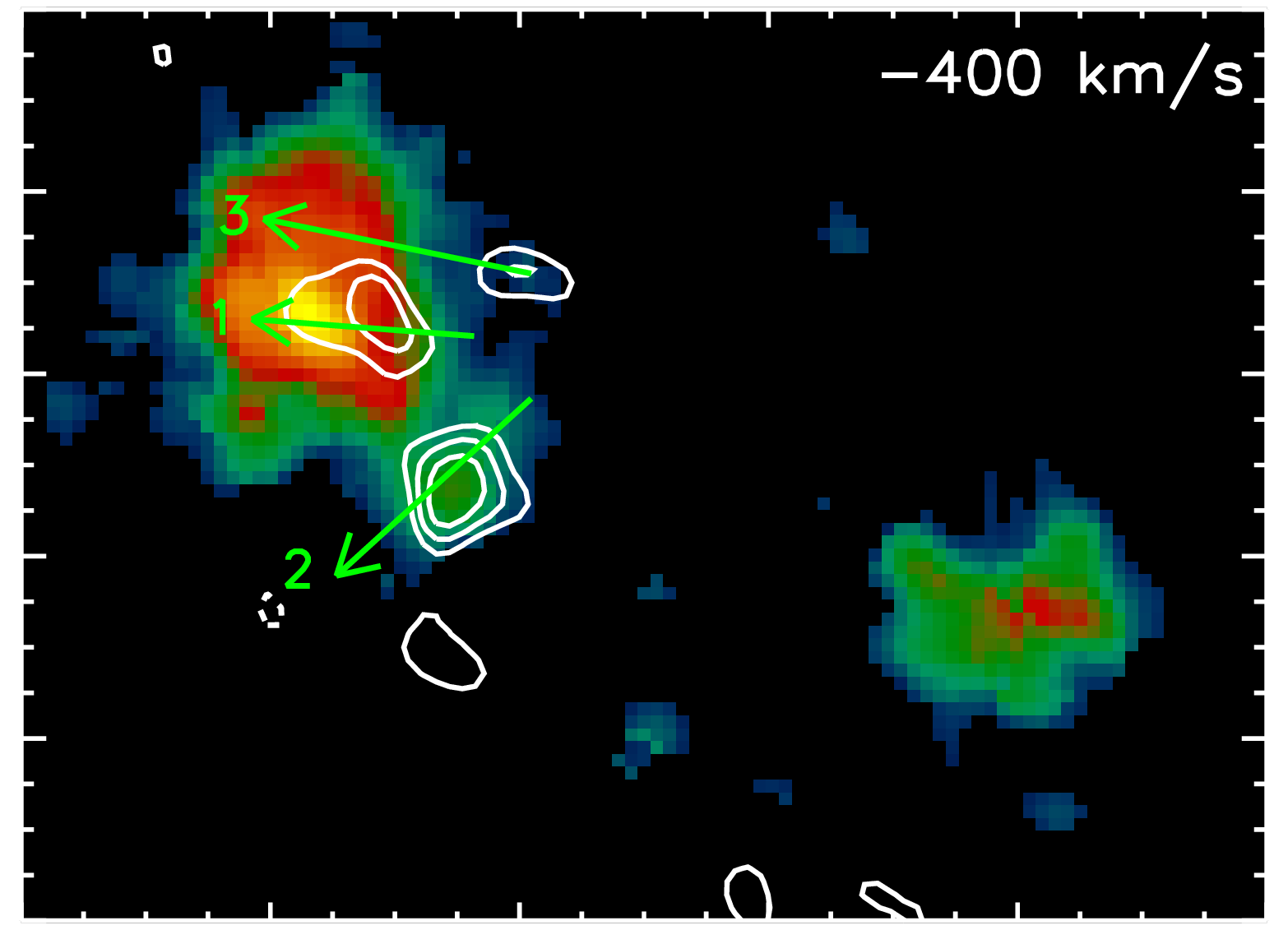} 
\hspace{-2.4mm}
\includegraphics[width=0.32\textwidth]{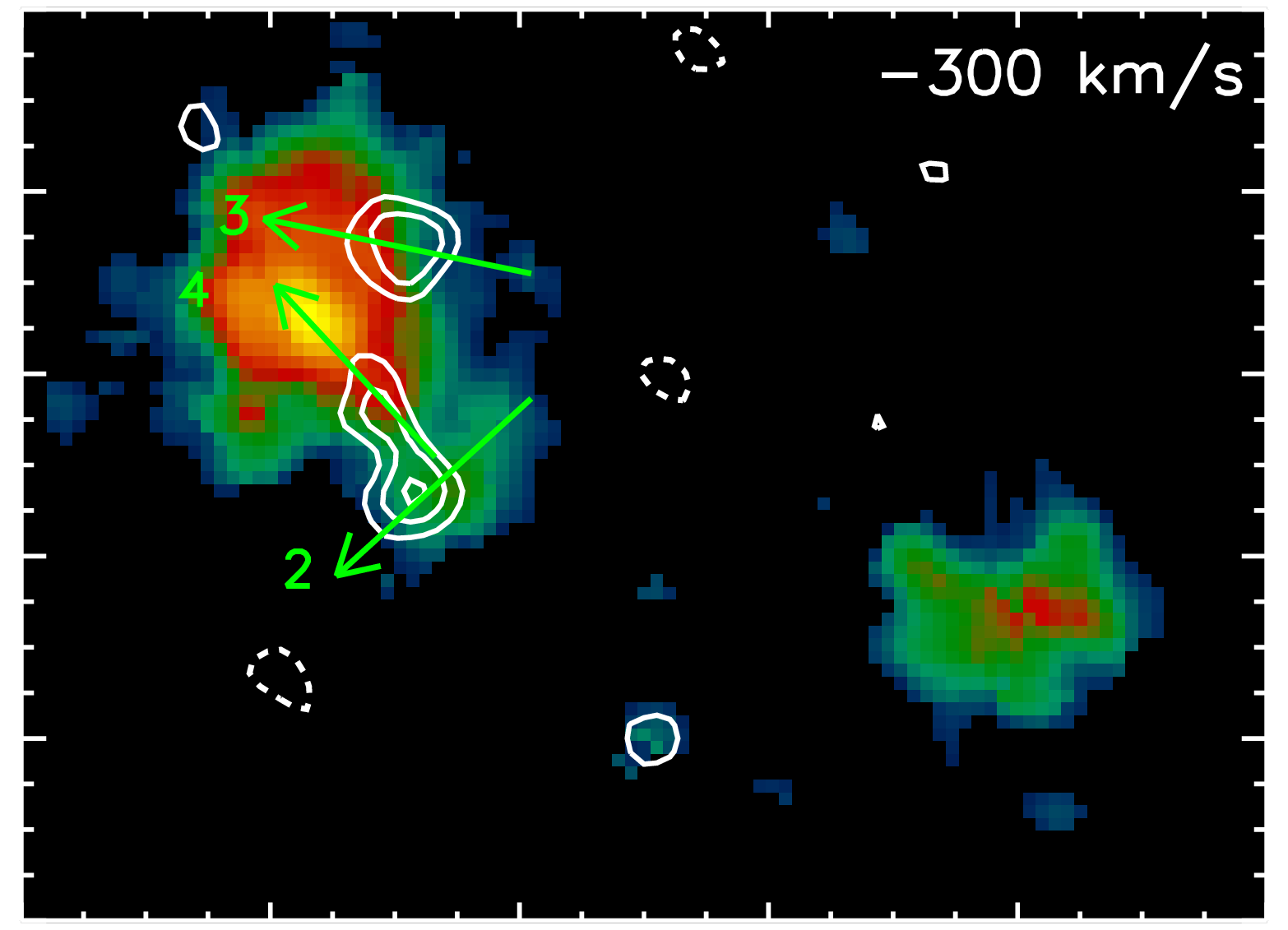} \\
\vspace{-0.5mm}
\includegraphics[width=0.32\textwidth]{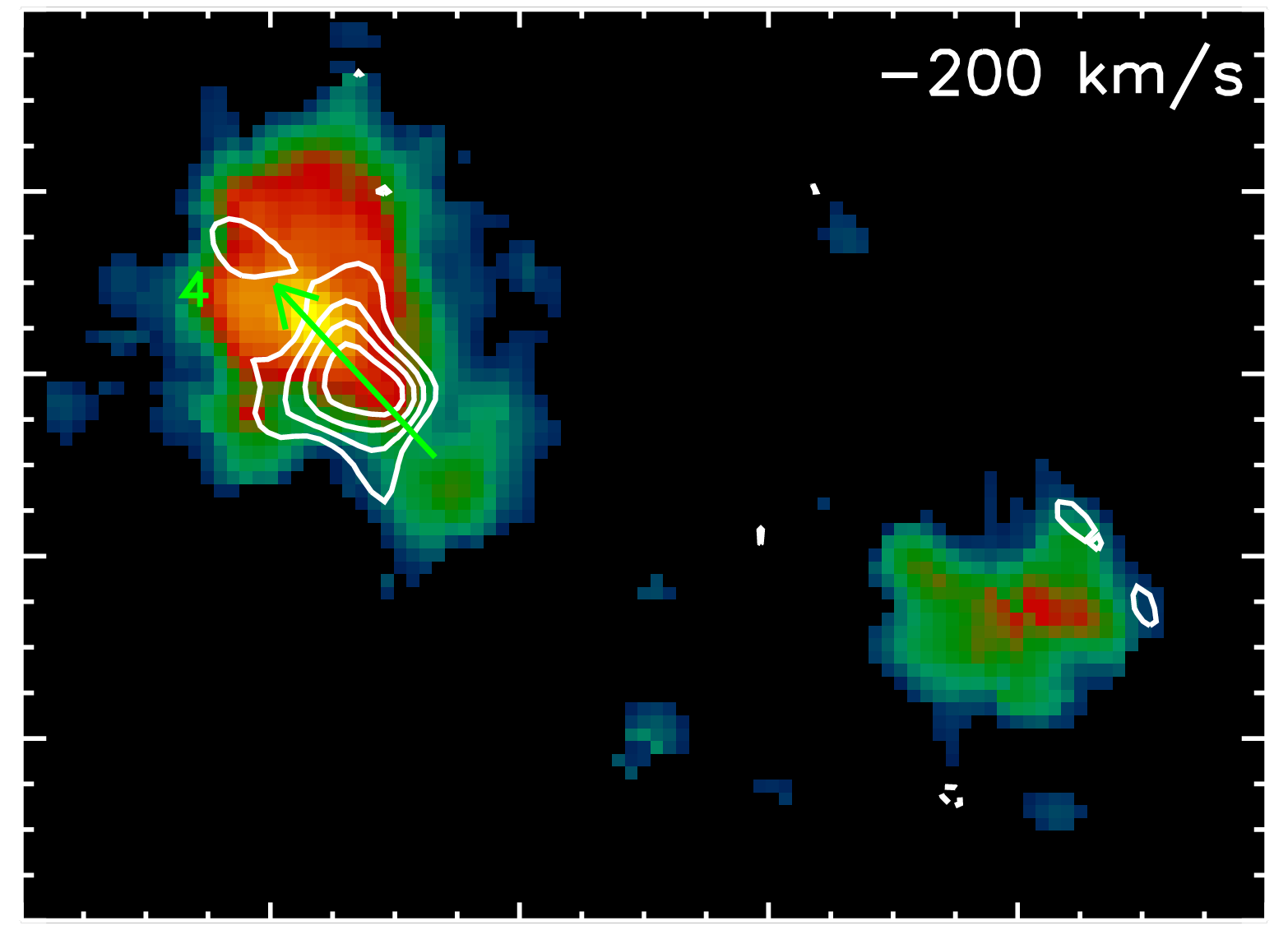} 
\hspace{-2.4mm}
\includegraphics[width=0.32\textwidth]{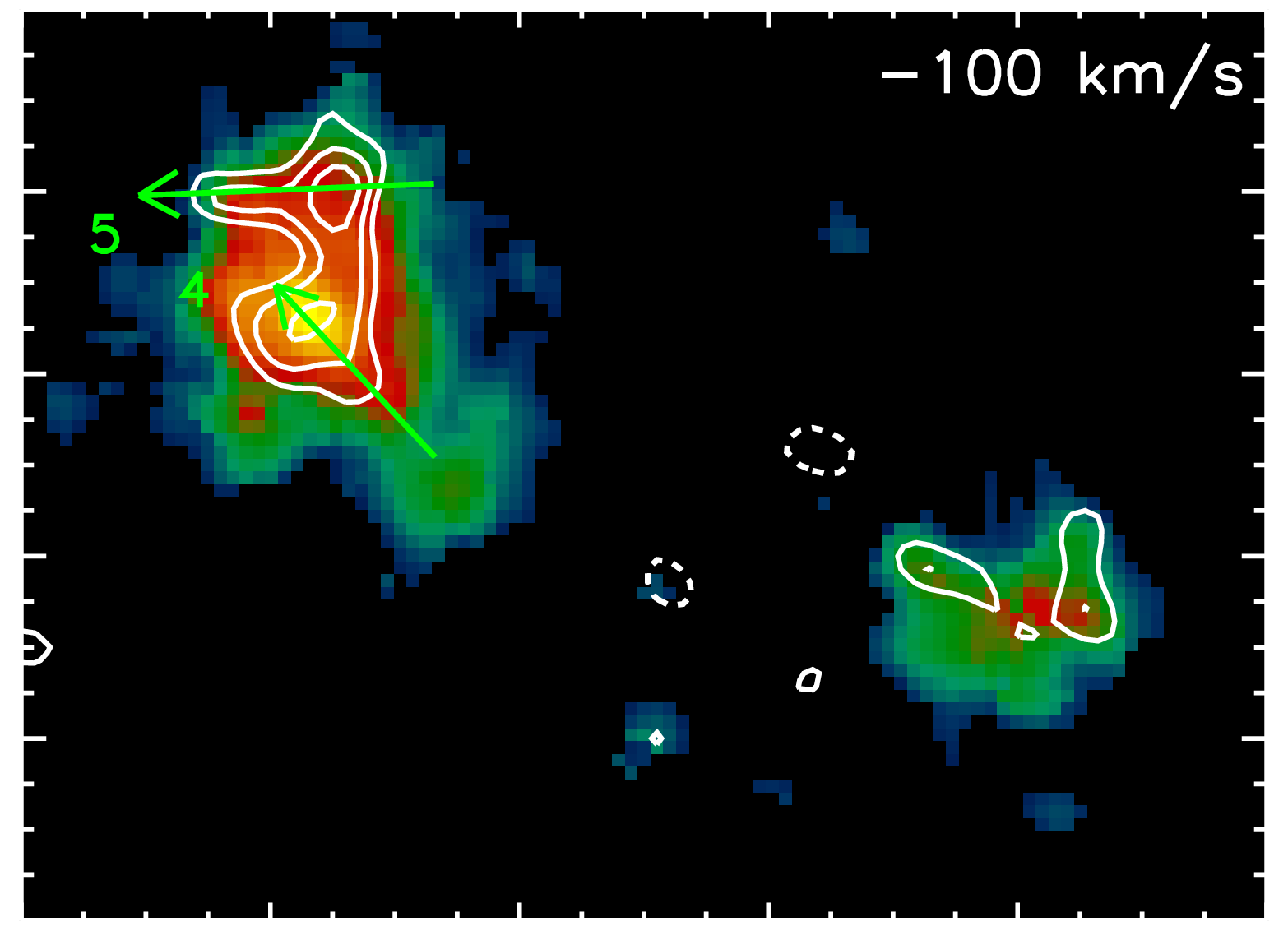} 
\hspace{-2.4mm}
\includegraphics[width=0.32\textwidth]{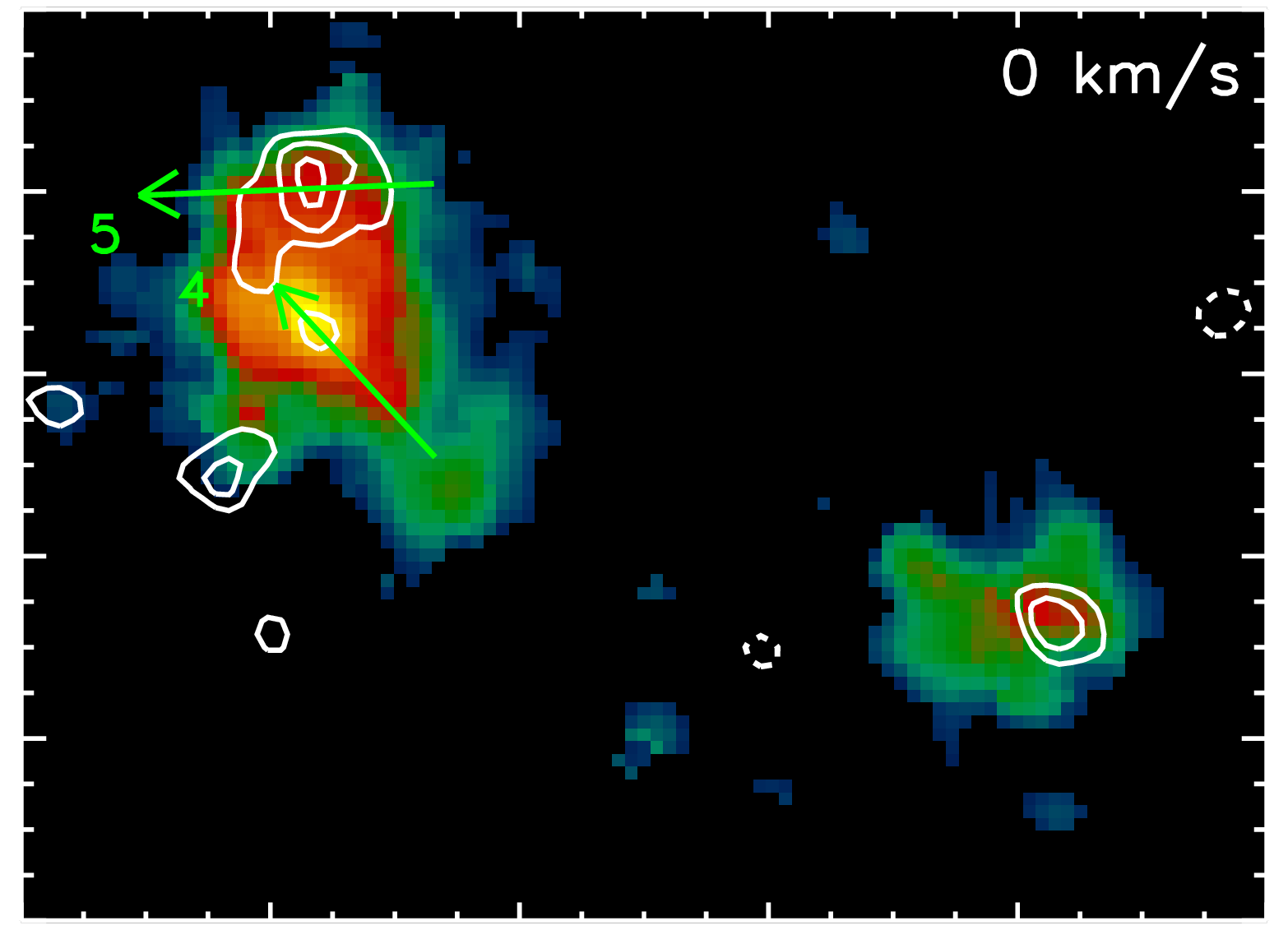}\\
\vspace{-0.5mm}
\includegraphics[width=0.32\textwidth]{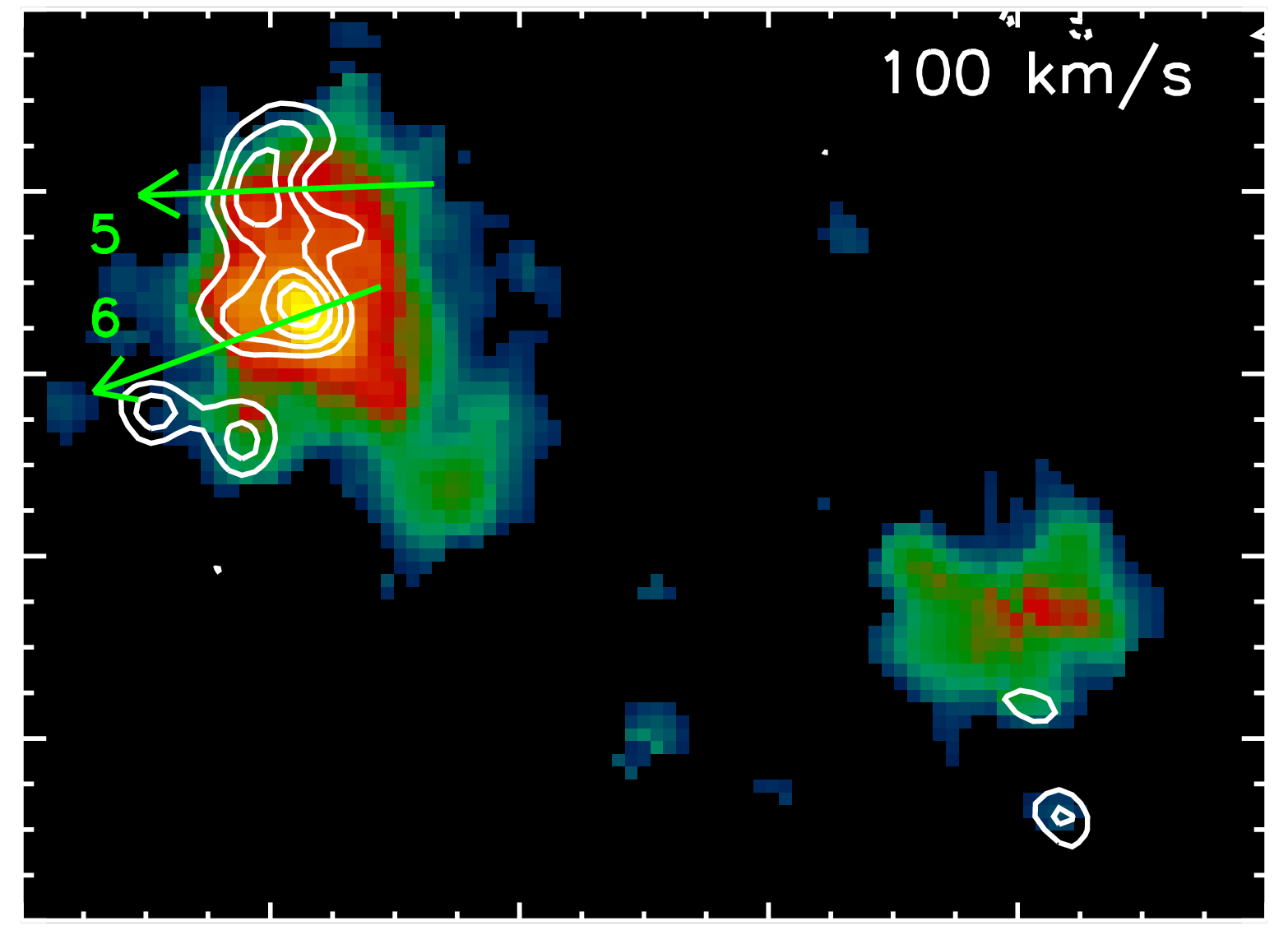} 
\hspace{-2.4mm}
\includegraphics[width=0.32\textwidth]{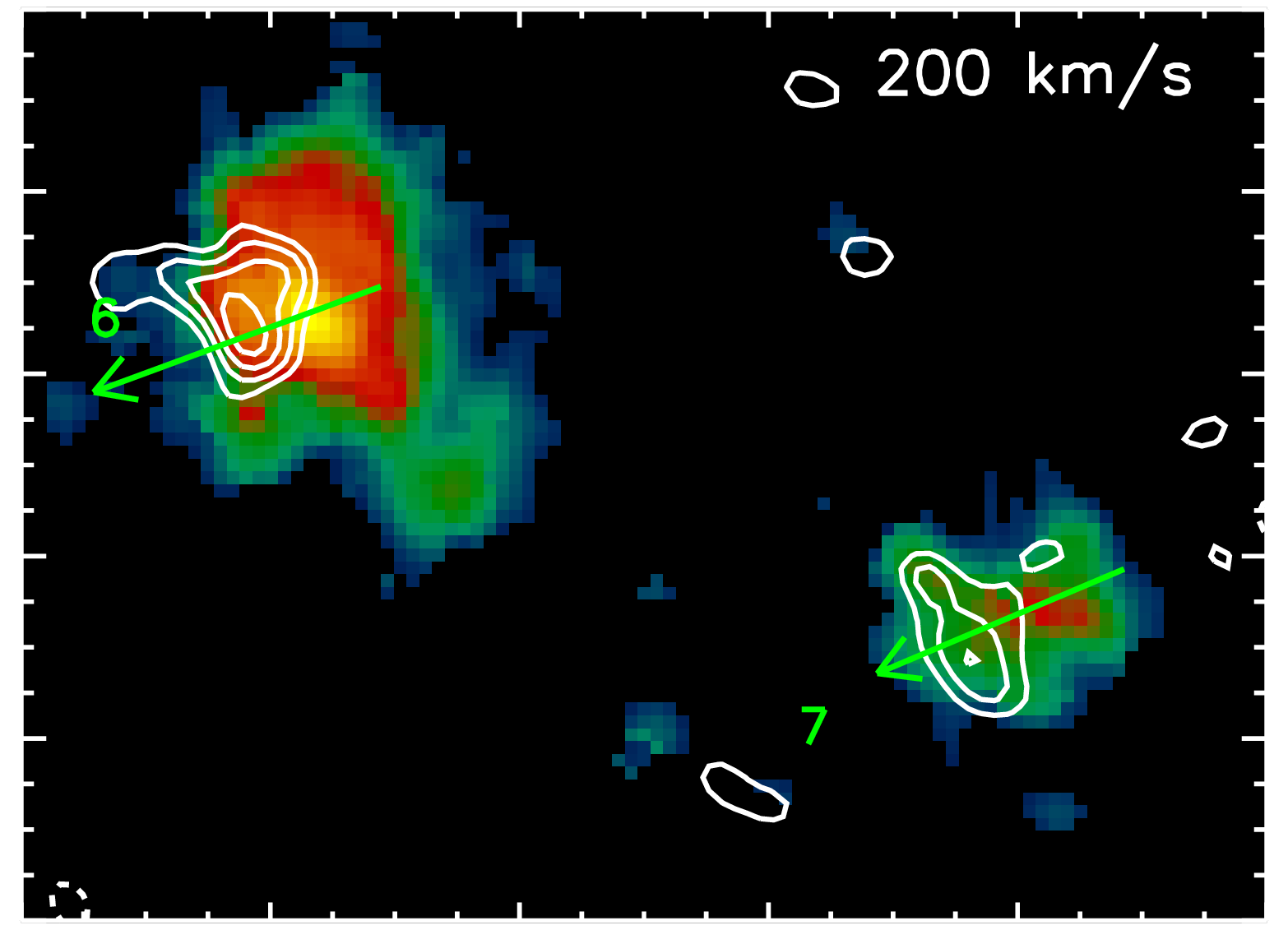} 
\hspace{-2.4mm}
\includegraphics[width=0.32\textwidth]{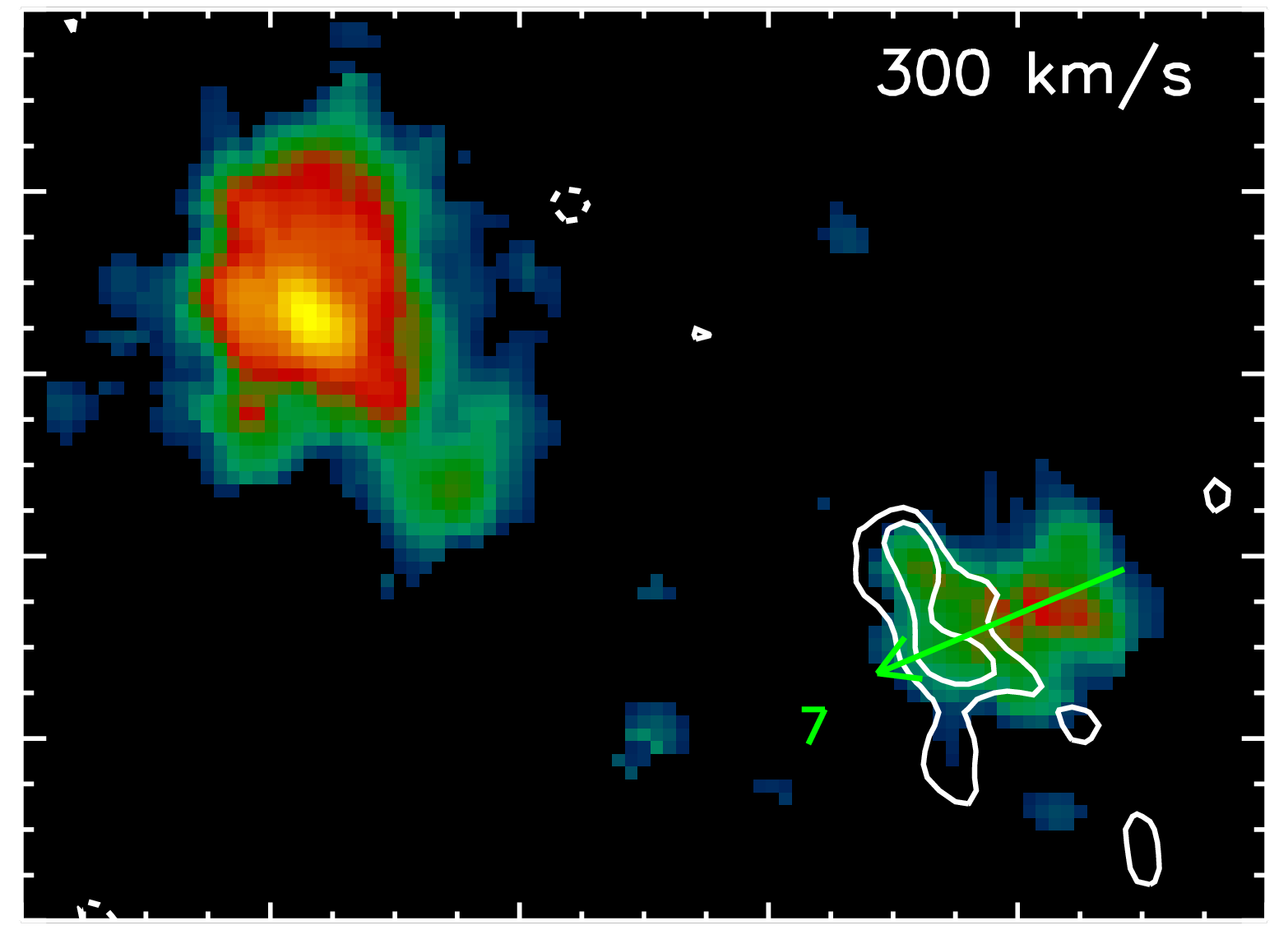} 
\caption{Dynamics of SGP38326 at high-spatial resolution from [CII] emission: [CII] velocity channel contours are overlaid on the [CII] moment-0 map. The velocity channels have been averaged over 100 ${\rm km \, s^{-1}}$. Contours in the velocity channels start at $\pm 3\sigma$ and are represented in steps of 1$\sigma$ (1$\sigma = 0.66 \, {\rm mJy \, beam^{-1}}$). Green arrows and associated numbers indicate the velocity gradients identified in the continuum-subtracted [CII] data cube.
              }
\label{velocity_map_CII_channels}
\end{figure*}


\subsection{Dust continuum morphology}\label{dust_morph_size}

Although SGP38326 is resolved into two interacting galaxies, each of
them is an extremely luminous starburst in its own right. The dust
continuum emission of SMG1 and SMG2 (rest-frame 160 $\mu$m) is
relatively smooth (see grey contours in
Fig.~\ref{radial_profiles_dust_CII_figure}). At this spatial
resolution, there is no clear evidence of clumpy emission in SMG1 and
SMG2. The physical extent of the dust emission has
been measured from a two-dimensional elliptical gaussian profile
fit. Table \ref{table_values_individual_sources} shows the FWHM of the
major and minor axis of each of the components of SGP38326. These values are
at the high-end of the size distribution of lower-redshift SMGs
\citep{Simpson2015ApJ...799...81S} and compatible with the size of $z > 3$ SMGs in \cite{Ikarashi2015ApJ...810..133I}. SMG1
and SMG2 are, however, considerably smaller than the dust emission in
GN20: $\sim 5.3\, {\rm kpc} \times 2.3 \, {\rm kpc}$
\citep{Hodge2015ApJ...798L..18H}.

Their size in combination with their IR-derived SFR (see Table
\ref{table_values_individual_sources}) imply that SMG1 and SMG2 have
median SFR surface densities of $\Sigma_{\rm SFR} \sim 840\, {\rm
  M_\odot \, yr^{-1} \, kpc^{-2}}$ and $\Sigma_{\rm SFR} \sim 570\,
{\rm M_\odot \, yr^{-1} \, kpc^{-2}}$, respectively. The $\Sigma_{\rm
  SFR}$ of SMG1 is very close to the theoretically predicted Eddington
limit for starburst disks that are supported by radiation pressure
\citep{Andrews2011ApJ...727...97A,Simpson2015ApJ...807..128S}. Such
high $\Sigma_{\rm SFR}$ values have been only found in the $z = 6.42$
quasar host galaxy J1148+5251 \citep{Walter2009Natur.457..699W} and
the dusty starbursts HFLS3 at $z = 6.34$ \citep{Riechers2013Natur.496..329R} and AzTEC-3 at $z \sim 5.3$
\citep{Riechers2014ApJ...796...84R} and
are considerably higher than the $\Sigma_{\rm SFR}$ found in other
starbursts at similar redshifts (see Table
\ref{table_comparison_sources}). 

Combining the sizes of the dust emission with the molecular gas mass estimates (assuming the molecular gas and dust are roughly co-spatial), we find that the estimated gas surface density of SMG1 and SMG2 are among the highest ever found (see Table \ref{table_values_individual_sources}). The values found for the SFR surface density, in combination with the derived gas mass surfaces densities place SMG1 and SMG2 between the sequence of starbursts and 'normal' star-forming galaxies in the $\Sigma_{\rm SFR}-\Sigma_{\rm H_2}$ plane, with gas depletion time scales between 10 and 100 Myr \citep{Hodge2015ApJ...798L..18H}. However, it should noted that this result assumes that the average CO line ratios for SMGs are valid for SGP38326.

\subsection{Dust properties and SED fitting}

The SPIRE fluxes for SMG1 and SMG2 are blended within the large SPIRE beam and the signal-noise ratio in the image is not high enough to attempt to deblend them using the ALMA 870$\mu$m positions. Instead we fit the global SED of both sources using both optically thin and optically thick modified black bodies to derive the dust temperatures and dust masses, as well as the integrated IR luminosity. To get the values for each source separately we have assumed they have the same SED shape and split the IR luminosity and dust mass in the ratio of the 870$\mu$m fluxes. Table \ref{table_values_individual_sources} lists the dust and SED properties. Because of the relatively low SNR of the SPIRE 250$\mu$m measurement and only upper limits for PACS, we have no strong preference for either an optically thin or optically thick model, both produce acceptable fits with comparable $\rm{L_{IR}}$. What differs strongly between the two models is the dust temperature and accordingly the dust mass which, for a rest
frame continuum measurement of 160$\mu$m, is strongly sensitive to the
assumed dust temperature. The optically thin model gives a best fit
temperature of $33\pm 2$ K, and dust masses of $\rm{M_{d(SMG1)} = 1.9
  \times 10^9 \, M_{\odot}}$ and $\rm{M_{d(SMG2)} = 6.9 \times 10^8 \,
  M_{\odot}}$, using a value of $\beta=2$ and
$\kappa_{850}=0.077\,{\rm m}^2/{\rm kg}^{-1}$
\citep{Dunne2001MNRAS.327..697D}. This dust temperature would be
remarkably low for such a highly star forming galaxy in the local
Universe but it is now well established that high-z SMGs have
much cooler SEDs for their infrared luminosity than do local ULIRGs,
due to the more widespread nature of their star formation \citep{Hwang2010MNRAS.409...75H,Elbaz2011A&A...533A.119E,Symeonidis2013MNRAS.431.2317S}.
The optically thick fit gives a higher dust temperature of 55 K with a
correspondingly lower dust mass of $5.6\times 10^8\,\rm{M_{\odot}}$
and $2.1\times 10^8\,\rm{M_{\odot}}$ for SMG1 and SMG2
respectively. The optical depth at 100$\mu$m is $\tau_{100} \sim 3$
in both galaxies. Dust is a good tracer of molecular gas at both high
and low redshift and is increasingly being used as an alternative to, or
consistency check for CO derived measurements \citep{Rowlands2014MNRAS.441.1017R,Scoville2015arXiv151105149S,Berta2015arXiv151105147B}. The derived gas mass and optically thin dust masses imply gas to dust ratios of $M_{\rm g}
/ M_{\rm d} \sim 101$ and $\sim 110$ for SMG1 and SMG2,
respectively. These values are compatible with those found for
starbursts at similar redshifts \citep{Magdis2011ApJ...740L..15M}, $z
\sim 0.5 - 1.5$ main-sequence galaxies and classical SMGs
\citep{Magdis2012ApJ...760....6M}, local star forming spirals
\citep{Leroy2011ApJ...737...12L, Sandstrom2013ApJ...777....5S} and local LIRGs and ULIRGs
\citep{Solomon1997ApJ...478..144S,Dunne2001MNRAS.327..697D}.

If we instead use the optically thick mass estimates we find higher
gas-to-dust ratios by a factor $\sim 3$, but in the extreme conditions of
these starbursts it would not be surprising to find both optically
thick dust SEDs {\em and} highly excited molecular gas, both of which
would lower the estimated molecular gas content by a factor of a few.

\begin{figure}[!t]
\centering
\includegraphics[width=0.44\textwidth]{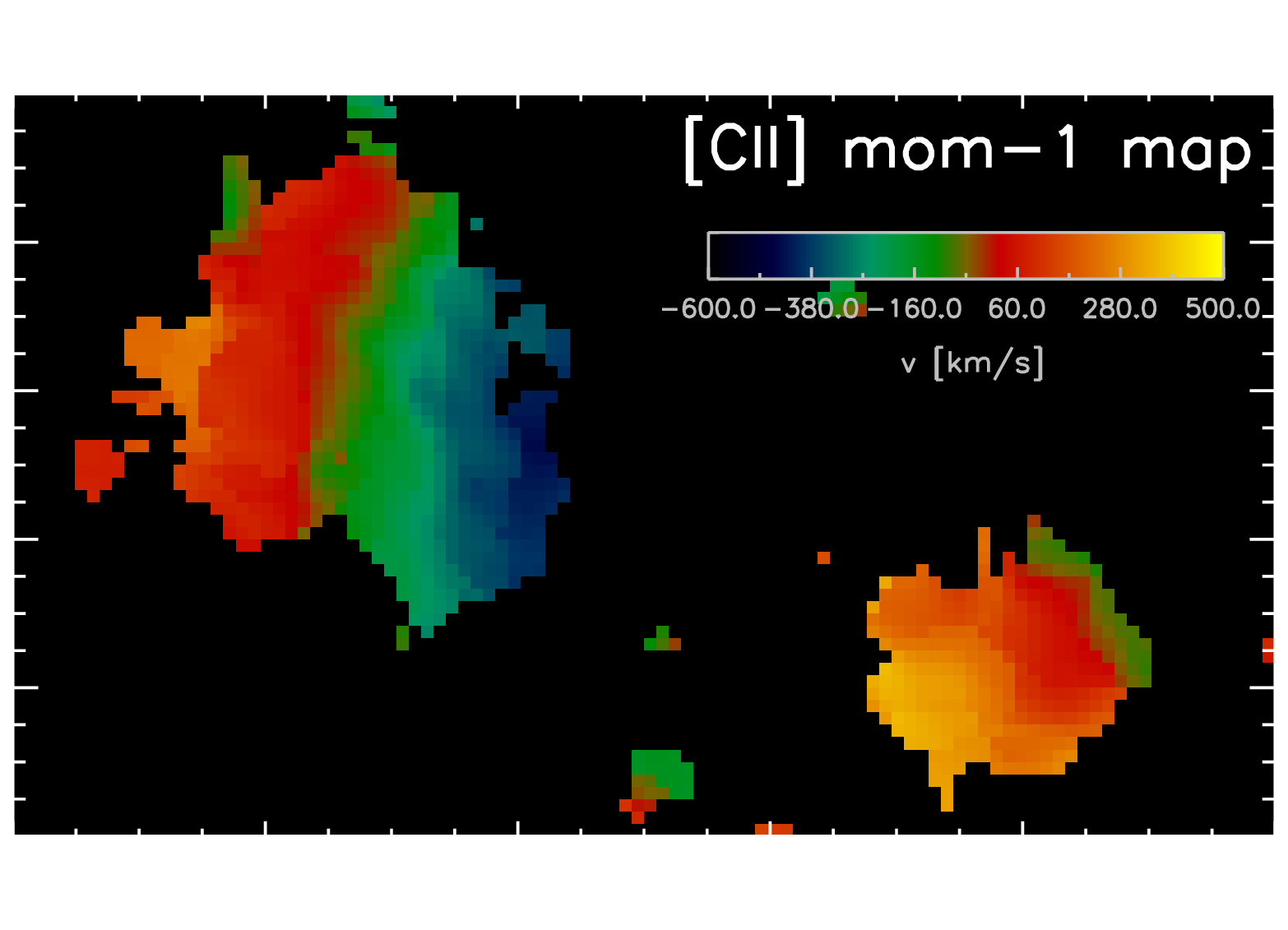} \\
\vspace{-11mm}
\includegraphics[width=0.44\textwidth]{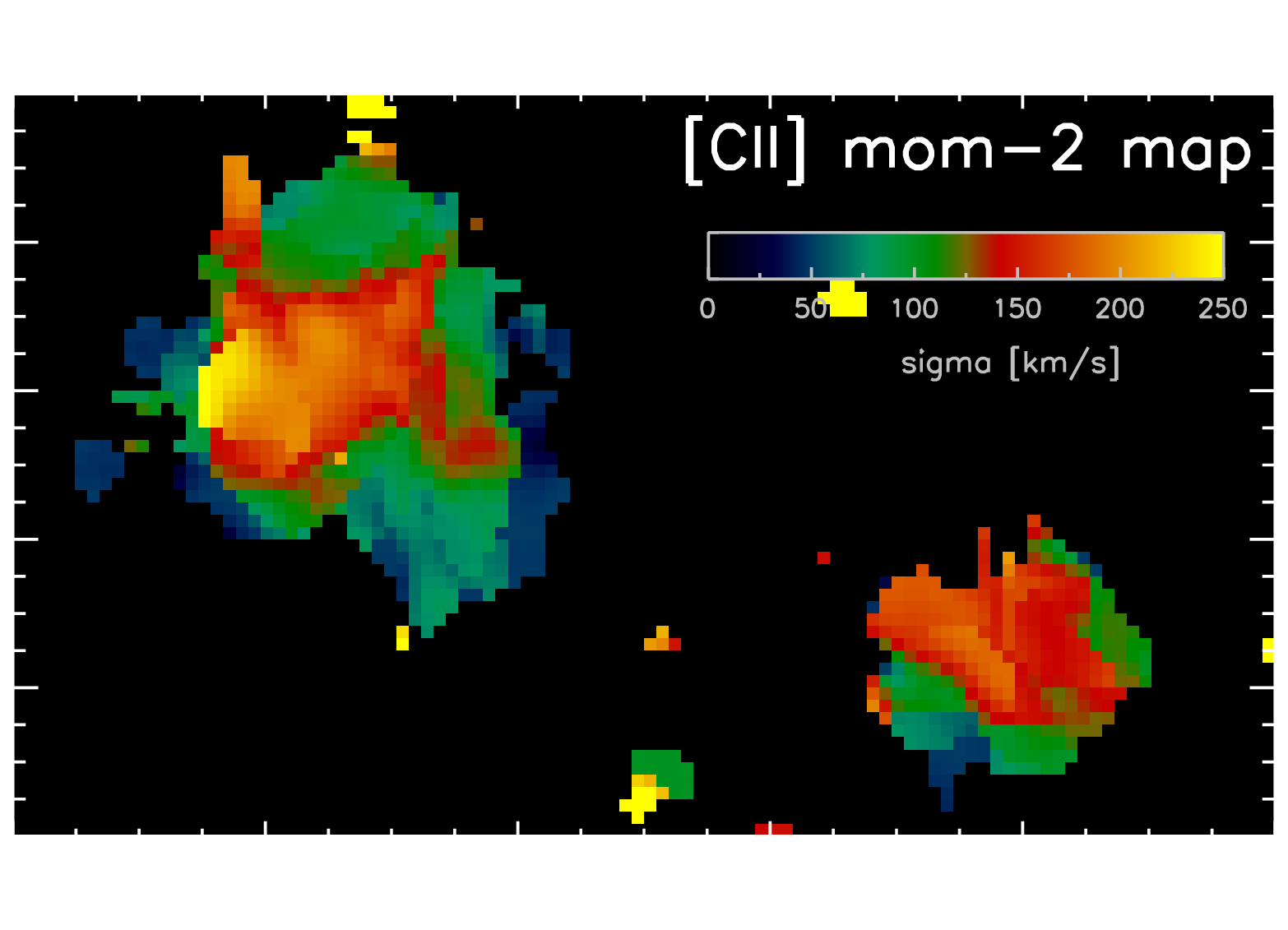}
\caption{Resolved velocity (\emph{upper}) and velocity dispersion (\emph{bottom}) maps of the two interacting components in SGP38326 obtained from their [CII] emission by using moment masking \citep{Dame2011arXiv1101.1499D}. As Figure \ref{velocity_map_CII_channels} suggests, all velocity gradients found in SMG1 and SMG2 are compatible with ordered rotation although with some perturbations, likely due to the effect of the ongoing merger.
              }
\label{fig_velo_moments12}
\end{figure}


\subsection{Morphology of [CII] emission}\label{CII_morphology}

The redshift of SGP38326 was unknown when the 870$\mu$m continuum ALMA
observations were performed. Fortunately, its redshifted [CII]
emission was covered by the default continuum setup, offering a unique
opportunity to study the ISM of intense starbursts at $z > 4$ at
unprecedented spatial resolution.

The [CII] spectrum of the two merging components of SGP38326 is
shown in Figure \ref{spec_CO_CII_lines}. SMG1 has a broad [CII] line
(${\rm 626 \pm 60 km \, s^{-1}}$ FWHM, $L_{\rm [CII]}/L_\odot = (8.3
\pm 0.2) \times 10^9$), considerably broader than the [CII] emission
in other SMGs at similar redshifts
\citep{Swinbank2012MNRAS.427.1066S,Walter2012Natur.486..233W,DeBreuck2014A&A...565A..59D,Riechers2014ApJ...796...84R}
and also broader than local ULIRGs
\citep{Farrah2013ApJ...776...38F}. The [CII] emission from SMG2 is
slightly narrower (${\rm 585 \pm 95 \, km \, s^{-1}}$ FWHM, $L_{\rm
  [CII]}/L_\odot = (2.9 \pm 0.2) \times 10^9$), with its center
shifted $\sim 300 \, {\rm km \, s^{-1}}$ with respect to SMG1. The
[CII] emission in SMG1 and SMG2 implies that their redshifts are
$z_{\rm SMG1} = 4.4237 \pm 0.0004$ and $z_{\rm SMG2} = 4.4289 \pm
0.0004$, respectively. The escape velocity of the system (assuming a
mass at least as high as the molecular gas mass) is significantly
greater than the velocity offset between SMG1 and SMG2, meaning that
the galaxies are destined to merge, in contrast to multiple
HyLIRGs seen at lower redshifts
\citep{Ivison2013ApJ...772..137I}. Therefore, our results support a
scenario where at least a subset of the most distant starbursts formed in the
same highly dissipative mergers of gas-rich galaxies that seem
ubiquitous amongst SMGs at $z < 3$ and local ULIRGs
\citep{Engel2010ApJ...724..233E}. This result is an agreement with \citep{Narayanan2015Natur.525..496N}, which reported that major mergers are common at $z > 4$ and indeed drive some short-lived burst in star formation.

The smooth dust morphology contrasts with the more irregular [CII]
emission (Fig.~\ref{radial_profiles_dust_CII_figure}). [CII] can be
associated with many ISM phases, such as photodissociation regions
associated with star formation \citep{Stacey1999ESASP.427..973S}, the
diffuse or dense molecular medium \citep{Wolfire2010ApJ...716.1191W}
and atomic and ionized gas
\citep{Madden1997ApJ...483..200M}. Additionally, [CII] can be enhanced
by shocks \citep{Appleton2013ApJ...777...66A} and altered by outflows
\citep{Cicone2015A&A...574A..14C}. In the [CII] velocity maps shown in
Figure \ref{velocity_map_CII_channels} there is no evidence for atomic
outflows. Therefore,
shock enhancement, local star formation conditions and tidal forces
triggered by the interaction between the components, coupled with
local excitation differences are the most likely explanations of the
different morphologies of [CII] and dust.

Figure \ref{radial_profiles_dust_CII_figure} also shows the profiles
of the [CII] and dust emission along the major and minor axis of the
[CII] emission of each starburst. In addition to their different
morphology, the gas traced by [CII] is extended over larger physical
scales than the dust
\citep{Tacconi2006ApJ...640..228T,Ivison2011MNRAS.412.1913I} and also
has significantly different radial profiles. This is further supported
by the same 2D gaussian fitting used to determine the size of the dust
emission. The size of the [CII] emission is $3.8 \pm 0.1 \, {\rm kpc}
\times 2.9 \pm 0.1 \, {\rm kpc}$ in SMG1 and $2.7 \pm 0.1 \, {\rm kpc}
\times 2.1 \pm 0.1 \, {\rm kpc}$ in SMG2 (FWHM values are quoted here). Therefore, we find that [CII] is extended over a spatial scale larger than that found for
dust (\S \ref{dust_morph_size}). This is in agreement to what it has been reported in other $z >
4$ sources, such as AzTEC-3 \citep{Riechers2014ApJ...796...84R}, HFLS3
\citep{Riechers2013Natur.496..329R}, and W2246-0526
\citep{Tanio2015arXiv151104079D}, but at lower spatial resolution
compared with the extention of the emission. Interestingly, some
faint, tidal [CII] emission seems to be flowing from SMG1 to SMG2 as a
consequence of their interaction. If confirmed, this would be similar
to the CO and [CII] bridges that have been previously seen in other interacting starbursting systems
\citep{Fu2013Natur.498..338F,Carilli2013ApJ...763..120C}.


\subsection{OH 163$\mu$m emission}\label{lack_OH_emission}

The default continuum observations also covered the 163 $\mu$m
OH($^{2}\Pi_{1/2} J=3/2 \rightarrow 1/2$) doublet (whose components
are at $\nu_{\rm rest} = 1834.74735$ and $1834.74735 \, {\rm
  GHz}$). We note that the nearby $^{12}$CO(16--15) emission
line is out of the spectral range covered. We present in Figure
\ref{spec_CO_CII_lines} the OH spectrum of SMG1 extracted in the area
where dust emission is detected (brown spectrum). We also show (blue
spectrum) the OH spectrum of SMG1 extracted on the position where the
dust emission peaks and using an aperture the size of the synthesized
beam. Assuming the redshift determined by the CO lines, the two
components of the OH doublet should be located at the velocities
marked by the short vertical grey lines. Despite the clear
  detection of the continuum, no OH emission is detected at the
expected velocities. However, there is an emission line in the blue
spectrum shifted by $\sim 200 \, {\rm km \, s^{-1}}$ blue-ward that
might be associated with the OH doublet since there is no other
emission line that could lie at that frequency. An absorption line is
apparent in the spectrum of SMG1 $\sim 100 \, {\rm km \, s^{-1}}$
red-ward of the expected position of any OH emission. However, the
centers of the emission and absorption do not match with the
redshifted components of the OH doublet. After continuum subtraction,
\citet{Riechers2014ApJ...796...84R} derived for AzTEC-3 a [CII] peak flux of about
$18 \, {\rm mJy}$ and a [OH] peak flux of about $2 \, {\rm mJy}$. This means a ratio of ${\rm [CII]/OH \sim 10}$. The peak [CII] flux
in SMG1 is $\sim 20\,{\rm mJy}$ and, assuming the same ${\rm [CII]/OH}$ ratio as in AzTEC-3, we would expect the OH emission to have a peak line flux of $\sim
2\,{\rm mJy}$, in agreement with the peak flux of the emission line
seen in the bottom panel of Figure \ref{spec_CO_CII_lines}. If the
detected emission line is actually one of the OH components, its
width ($\sim 85 \, {\rm km} \, {\rm s}^{-1}$) is lower than that
previously found in strong starbursts in the distant Universe
\citep{Riechers2013Natur.496..329R,Riechers2014ApJ...796...84R}. We
note that these are the only two cases where OH 163$\mu$m emission has
been clearly detected in $z > 4$ starbursts. The putative OH
emission in SMG1 is also very compact, being spatially coincident with
the maximum of the dust continuum emission. It should be pointed out
that the OH emission is only seen where the dust emission is
maximum. In contrast, the absorption is seen across the whole disk
(brown spectrum in Figure \ref{spec_CO_CII_lines}).


\subsection{Dynamics at high resolution from [CII] emission}\label{dynamics_from_CII_line}

The signal-to-noise of the [CII] detection allows us to study the
dynamics of SMG1 and SMG2 to a level of detail never previously
achieved for any $z > 4$ unlensed starburst. We present in Figure
\ref{velocity_map_CII_channels} the [CII] emission in all the velocity
channels where it is detected. We have identified six different
velocity gradients in SMG1 and one in SMG2, which reveal that both
components of SGP38326 are ordered rotating disks, as corroborated by
the [CII] velocity and velocity dispersion maps shown in Figure
\ref{fig_velo_moments12}. It should be noted that the velocity field
derived in SMG1 from [CII] is very similar to that obtained from the
$^{12}$CO(5--4) transition (see Figure \ref{dust_CO_morph_3comp}), but
with about 10x times better spatial resolution. We have fitted the
observed velocity field using a disk plus dark matter halo model. We
should note that, since the rotation curve does not turn over, the
disk+halo model is degenerate. In any case, if it is a settled
disk, it must be massive, with $v_{\rm c} \sim 300 \, {\rm km \,
  s^{-1}}$ and with a best-fit inclination of $\sim 65 \, {\rm
  deg}$. This suggests a dynamical mass $M_{\rm dyn,\, SMG1} \sim 5
\times 10^{10} \, M_\odot$.  
The dynamical analysis from [CII] emission
at high-spatial resolution indicates that the wide, unresolved CO
emissions in SGP38326 (see Figure \ref{spec_CO_CII_lines}) are driven
by the interaction of two disks separated by $\sim 200\, {\rm km/s}$
in velocity space. This is opposite to, for example, to AzTEC-3, where
the [CII] emission (which has half the width of the [CII] emission in
SMG1) is supported by emission from a highly dispersed gas
\citep{Riechers2014ApJ...796...84R}.

It should be pointed out that the derived dynamical mass is $\sim
  \times 5$ lower than the molecular gas mass estimated either
  from CO or the optically thin dust SED fit. However, there are
  considerable uncertainties involved in the calculation of both of
  these quantities. The dynamical mass estimate is uncertain by
at least by a factor of $\sim$2 due to the uncertainty in the
inclination of the disk and the merger state. The estimation of the
molecular gas mass assumes an $\alpha_{\rm CO}$ factor (we used the value typically
assumed for ULIRGs, see above) and a ratio between the $^{12}$CO(5--4)
and $^{12}$CO(1--0) transitions. The derived molecular gas mass would
be lower than the obtained value if $\alpha_{\rm CO}$ is actually lower or
if the mid-$J$ CO lines are more excited than the average SMG.

We have also determined the dynamical mass of SMG1 and SMG2 by using
the isotropic virial estimator
\citep{Tacconi2008ApJ...680..246T,Forster2009ApJ...706.1364F,Engel2010ApJ...724..233E}:

\begin{equation}
M_{\rm dyn, \, vir} = 2.8 \times 10^5 \Delta v_{\rm FWHM}^2 R_{\rm 1/2}
\end{equation}

\noindent where $\Delta v_{\rm FWHM}$ is the width of the integrated
          [CII] line of each component and $R_{\rm 1/2}$ is the radius
          of the size of the [CII] emission. Since the size of the
          [CII] emission in SMG1 and SMG2 was derived by assuming
          elliptical gaussian profiles (see \S \ref{CII_morphology}),
          $R_{\rm 1/2}$ is assumed to be the average value of the semi
          major and minor axis (measured from FWHM). The derived
          dynamical masses with this method for SMG1 and SMG2 ($M_{\rm
            dyn}^{\rm SMG1} \sim 3.6 \times 10^{11} M_\odot$ and
          $M_{\rm dyn}^{\rm SMG2} \sim 2.3 \times 10^{11} M_\odot$)
          are now higher than the molecular gas masses. As pointed
          out in \cite{Engel2010ApJ...724..233E}, the scaling factor
          appropriate for a rotating disk at an average inclination is
          a factor of $\sim 1.5$ smaller
          \citep{Bothwell2010MNRAS.405..219B}. Applying this
          correction to our measurements, the derived dynamical masses
          would still be higher than the molecular gas masses. In any
          case, to carry out a meaningful comparison between the
          molecular gas and dynamical mass and put constraints on the
          $\alpha_{\rm CO}$ conversion factor, low-$J$ CO observations
          would be needed.

We now examine whether the two disks in SGP38326 are stable by using
the Toomre parameter $Q$ \citep{Toomre1964ApJ...139.1217T},
  which characterizes the stability of a disk against gravitational
fragmentation. In this sense, a disk is stable if $Q > 1$. We
calculate the Toomre parameter by following
\cite{Swinbank2015ApJ...806L..17S}: $Q = \sigma_r \kappa / \pi G
\Sigma_{\rm gas}$, where $\kappa = a V_{\rm max} / R$ is the epicyclic
frequency (with $a = \sqrt{3}$), $\sigma_r$ is the radial
velocity dispersion, and $\Sigma_{\rm gas}$ is the gas mass surface
density. All the dynamical parameters are obtained from the [CII] emission. We obtain $Q^{\rm SMG1} \sim 0.22$
and $Q^{\rm SMG2} \sim 0.35$, indicating that the two disks are unstable. For comparison, \cite{Swinbank2015ApJ...806L..17S}
obtained $Q = 0.30 \pm 0.10$ for SDP.81, a strongly lensed galaxy at
$z \sim 3$ observed with ALMA at very high spatial resolution
\citep{Vlahakis2015ApJ...808L...4A,Dye2015MNRAS.452.2258D}, while \cite{DeBreuck2014A&A...565A..59D} obtained an average $Q = 0.58 \pm 0.15$ for a $z \sim 4.8$ starburst containing an AGN. It should be noted that the derived value of the Toomre parameters relies on the validity for SGP38326 of the average CO line ratio for SMGs. This uncertainty is also present in previous works \citep{Swinbank2015ApJ...806L..17S,DeBreuck2014A&A...565A..59D}.


\subsection{The spatially resolved $L_{\rm [CII]} / L_{\rm IR}$ ratio}

\begin{figure}[!t]
\centering
\includegraphics[width=0.44\textwidth]{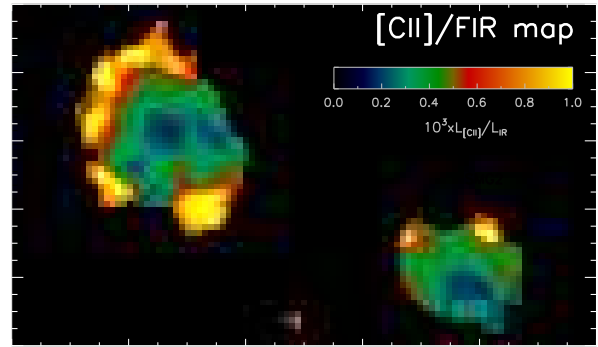}
\caption{Resolved $L_{\rm [CII]} / L_{\rm IR}$ ratio map for our pair of SMGs. The $L_{\rm IR}$ for each pixel has been calculated from the observed flux density and using an extrapolation based on the median SED of SMGs in the ALESS survey \citep{Swinbank2014MNRAS.438.1267S}, which produces a good fit to our photometric data points. The different morphology of dust and [CII] means that the [CII]/FIR ratio is higher in the outer parts of the galaxies and lower in the centre (where the peak of dust continuum emission is located). It should be noted that the yellow regions actually represent a lower limit of the $L_{\rm [CII]} / L_{\rm IR}$ ratio, since they correspond to regions where dust emission is not detected at $> 3\sigma$.
              }
\label{cii_dust_ratio_map}
\end{figure}

\begin{figure*}[!t]
\centering
\includegraphics[width=0.52\textwidth]{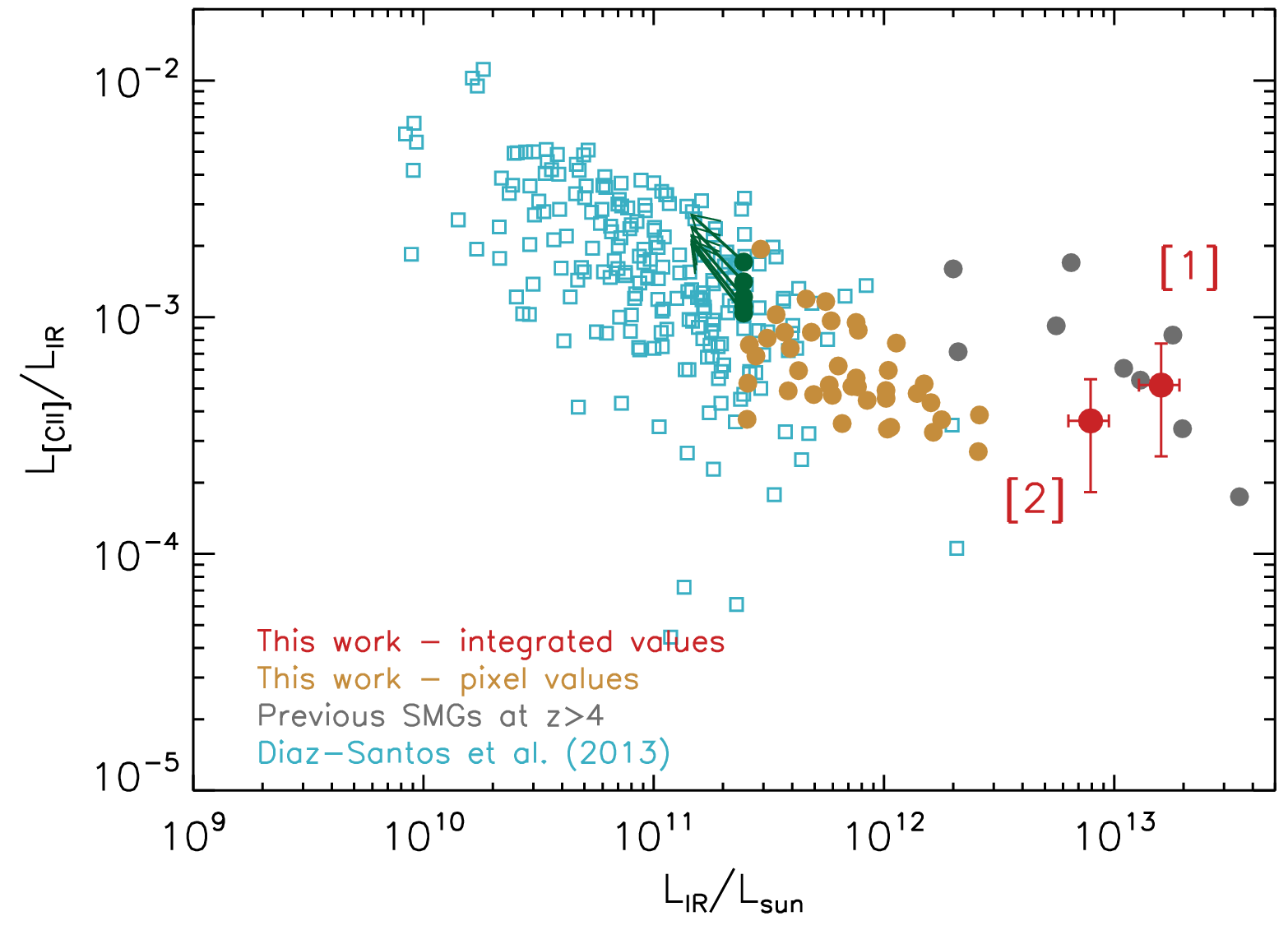}
\hspace{-15mm}
\includegraphics[width=0.52\textwidth]{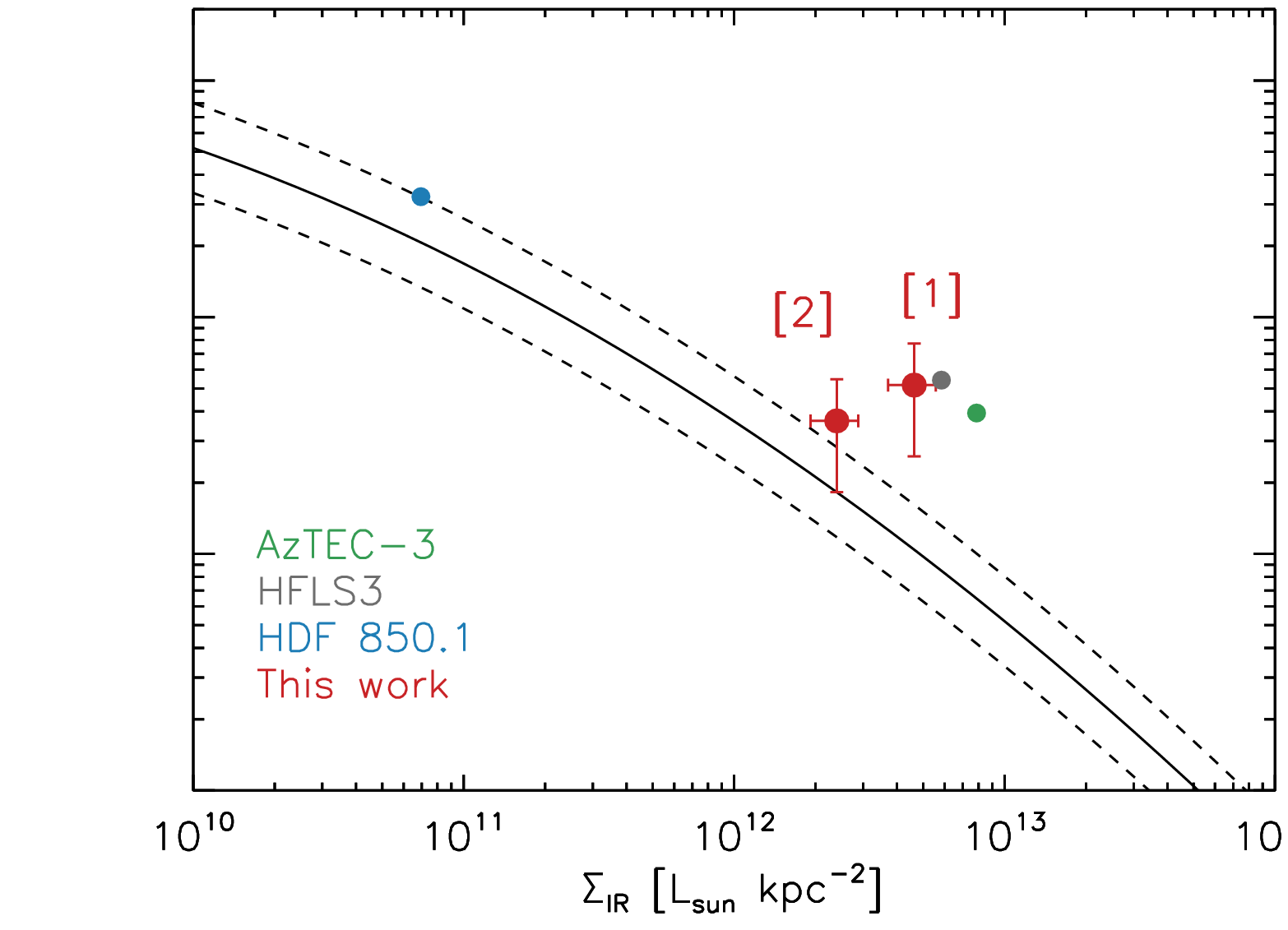}
\caption{[CII] to $L_{\rm IR}$ ratio for our pair of interacting starbursts as a function of their total IR luminosity (\emph{left}) and their IR surface density (\emph{right}). On the left panel, the values for the two SMGs (SMG1 being [1] and SMG2 being [2]) in SGP38326 are shown with red filled dots and compared with previous results (grey dots) for submm galaxies at similar or higher redshifts \citep{Swinbank2012MNRAS.427.1066S,Walter2012Natur.486..233W,Riechers2013Natur.496..329R,Rawle2014ApJ...783...59R,DeBreuck2014A&A...565A..59D,Riechers2014ApJ...796...84R,Tanio2015arXiv151104079D} and with local IR-bright galaxies \citep{DiazSantos2013ApJ...774...68D}. For lensed sources, the values included in the Figure has been corrected for amplification. We also represent the individual pixel values on the $L_{\rm [CII]} / L_{\rm IR}$ map (brown dots) after resampling the pixels to make them match with the size of the synthesized beam (see text for details). Green dots with arrows indicate upper limits of the $L_{\rm [CII]} / L_{\rm IR}$ and correspond to regions where there is [CII] emission but no dust continuum detection at $3 \sigma$. The IR luminosities in \cite{DiazSantos2013ApJ...774...68D} have been converted to $L_{\rm 8 - 1000 \mu m}$ by multiplying by a 1.8 factor. Our SMGs are on the bright side of the brightness distribution of previous SMGs with [CII] detections, and they also have amongst the lowest $L_{\rm [CII]} / L_{\rm IR}$ ratios. In the right panel, the black curve represents the best-fit for local galaxies in \cite{Lutz2015arXiv151102075L}, with the dashed curves representing the scatter of the local relation. We compare the location of our two SMGs with dusty starbursts at similar or higher redshifts. Despite the small number of galaxies, there is an indication that galaxies with the highest SFR and star formation surface density (SMG1, AzTEC-3 and HFLS3) tend to depart from the local relation towards higher values of the $L_{\rm [CII]} / L_{\rm IR}$ ratio.
              }
\label{cii_dust_ratio_integrated_values}
\end{figure*}

Our high-spatial resolution observations also allow us to obtain the
first resolved $L_{\rm [CII]}/L_{\rm IR}$ ratio map in a dusty
starburst at $z > 4$. We have obtained the $L_{\rm [CII]}/L_{\rm IR}$
ratio map by simply dividing the $L_{\rm [CII]}$ and $L_{\rm IR}$
maps. It should be noted that no convolution is required since the
spatial resolution of the [CII] and dust continuum map is the
same. The $L_{\rm [CII]}$ map has been obtained from the [CII]
moment-0 map by converting the line intensities in each pixel
into line luminosities. The $L_{\rm IR}$ map has been derived from the
observed 870$\mu$m dust continuum map by assuming the conversion from
flux density to total IR luminosity at $z = 4.425$ derived from the
average FIR SED of SMGs in ALESS \citep{Swinbank2014MNRAS.438.1267S},
which is a good representation of the FIR SED of
SGP38326. 

The distinct dust and [CII] morphology leads to a complex spatial
distribution for the [CII] to $L_{\rm IR}$ ratio
(Fig.~\ref{cii_dust_ratio_map}), which is higher in the outer parts of
the two starbursts and lower in their centres, in agreement with what
is observed for local LIRGs \citep{Tanio2014ApJ...788L..17D}. It
should be noted that the yellow regions in Figure
\ref{cii_dust_ratio_map} indicate a lower limit for the $L_{\rm
  [CII]}/L_{\rm IR}$ ratio since they correspond to regions where dust
continuum has not been detected, but where there is detectable
[CII] emission (recall that [CII] is extended over a larger area
than the dust, \S \ref{CII_morphology}). Explaining the irregular
morphology of the $L_{\rm [CII]}/L_{\rm IR}$ ratio is challenging due
to the many phases of the ISM giving rise to [CII] emission (see
\S \ref{CII_morphology}). The most likely explanations are
different excitation conditions for [CII] and variable dust
  optical depth across the disks. The presence of an AGN can also
lower the $L_{\rm [CII]}/L_{\rm IR}$ ratio
\citep{DiazSantos2013ApJ...774...68D}, and so an AGN could
therefore be responsible for the low values of the $L_{\rm
  [CII]}/L_{\rm IR}$ ratio in the center of the galaxies and in their
integrated values (see below). While we have no direct evidence
  for an AGN with the present data, the existence of one would not be
surprising since AGN seem to be ubiquitous in the strongest
starbursts at lower redshifts \citep{Ivison2013ApJ...772..137I}. 

Larger samples of
galaxies would be needed to explore at high resolution whether these
gradients are present in most of the population of dusty
starbursts at $z > 4$ and also in less extreme star forming systems.

Figure \ref{cii_dust_ratio_integrated_values} compares the integrated
values of the $L_{\rm [CII]}/L_{\rm IR}$ ratio with sources at similar
redshifts and in the local Universe. The left panel of Figure
\ref{cii_dust_ratio_integrated_values} shows that the values found for
the $L_{\rm [CII]}/L_{\rm IR}$ ratio in SMG1 and SMG2 are similar
  to those found in other comparably luminous IR galaxies at similar and
higher redshifts
\citep{Swinbank2012MNRAS.427.1066S,Walter2012Natur.486..233W,DeBreuck2014A&A...565A..59D,Riechers2014ApJ...796...84R,Tanio2015arXiv151104079D}
and in the local Universe \citep{DiazSantos2013ApJ...774...68D}. Our
two interacting starbursts are at the upper end of the
$L_{\rm IR}$ distribution of $z > 4$ galaxies, suggesting that the
anti-correlation between $L_{\rm [CII]}/L_{\rm IR}$ ratio and $L_{\rm
  IR}$ seen in the local Universe holds up to $z \sim 4$, but with an
offset in $L_{\rm IR}$.

We also present in Figure \ref{cii_dust_ratio_integrated_values} the
individual pixel values of the $L_{\rm [CII]}/L_{\rm IR}$ ratio for
SMG1 and SMG2. Since the synthesized beam is sampled by approximately
5 pixels on each size, we have re-binned the $L_{\rm [CII]}/L_{\rm
  IR}$ ratio map every 5 pixels, so each new pixel approximately
samples the area of one synthesized beam. Average values across
contiguous pixels are taken during map re-sizing. For those regions
with [CII] but no dust continuum detection we have given lower limits
instead. The individual pixel values suggest that galaxy regions
associated with higher star formation (or higher dust surface density) have lower $L_{\rm [CII]}/L_{\rm
  IR}$ ratios.

A tight correlation between the $L_{\rm [CII]}/L_{\rm IR}$ ratio and
the IR surface density has been reported in local galaxies
\citep{Lutz2015arXiv151102075L}. This relation and its typical scatter
is represented in the right panel of Figure
\ref{cii_dust_ratio_integrated_values} along with the derived values
for SMG1 and SMG2 and other $z > 4$ starbursts with measured sizes. We
have defined the IR surface density as $\Sigma_{\rm IR} = L_{\rm IR} /
(\pi R_{\rm a} R_{\rm b})$, where $R_{\rm a}$ and $R_{\rm b}$ are the
major and minor axis radii (derived by ${\rm FWHM / 2}$ from a
elliptical gaussian profile) of the continuum dust emission of each
galaxy. 

Interestingly, the local relation holds, within the
uncertainties, for SMG2 and also for other sources at $z >
4$. However, SMG1 and the most extreme sources at high redshift such
as AzTEC-3 or HFLS3 depart from that relation, having $L_{\rm
  [CII]}/L_{\rm IR}$ ratios higher than the local relation for a given
IR surface density. SMG1, AzTEC-3 and HFLS3 are close to the Eddington
limit. This might indicate that the [CII] emission in those galaxies
might be excited by an outflow. In fact, strong OH 163$\mu$m emission
has been detected in AzTEC-3 and HFLS3
\citep{Riechers2013Natur.496..329R,Riechers2014ApJ...796...84R},
supporting the presence of outflowing material. However, no clear
signature of OH 163$\mu$m emission or outflowing material (apart from
the extended $^{12}$CO(5--4) emission shown in Figure
\ref{dust_CO_morph_3comp}) has been detected in SMG1 (see \S
\ref{lack_OH_emission}), leaving unknown the reason of its enhanced
$L_{\rm [CII]}/L_{\rm IR}$ ratio with respect to its IR surface
density.


\section{Conclusions}\label{concluuuuu}

In this work we have presented ALMA high-spatial resolution ($\sim 0.1''$ or $\sim 800 \,{\rm pc}$) imaging of the gas and dust in SGP38326, an interacting pair of IR-bright starburst at $z = 4.425$. Despite the redshift of the two galaxies being unknown when the ALMA 870$\mu$m continuum observations were defined, the [CII] and OH emission lines were luckily covered by the continuum spectral setup. This offered the opportunity of studying at high resolution the properties of the ISM in the likely progenitor of a massive, elliptical galaxy at $z \sim 3$. The main conclusions of our work are:

\begin{itemize}

\item The dust continuum emission (rest-frame 160$\mu$m) in our pair of interacting starbursts at $z = 4.425$ presents a relatively smooth distribution at the resolution of our data and is distributed over an area of $2.2 \pm 0.2 \,{\rm kpc} \times 2.0 \pm 0.2 \,{\rm kpc}$ in SMG1 and $2.1 \pm 0.2 \,{\rm kpc} \times 1.5 \pm 0.1 \,{\rm kpc}$ in SMG2 (values quote FWHM of a 2D gaussian elliptical profile). However, the gas traced by [CII] emission has a more irregular morphology and is more extended than the gas, being extended over an area of $3.8 \pm 0.1 \, {\rm kpc} \times 2.9 \pm 0.1 \, {\rm kpc}$ in SMG1 and $2.7 \pm 0.1 \, {\rm kpc} \times 2.1 \pm 0.1 \, {\rm kpc}$ in SMG2.

\item  A dynamical analysis of the resolved [CII] emission revealed that the velocity fields of the two interacting components of SGP38326 are compatible with disk-like ordered ration, in contrast with other extreme IR-bright starburst in the early Universe. The velocity field of the two interacting starbursts does not show any clear sign of outflowing material. In fact, the 163 $\mu$m emission line (a tracer of molecular outflows), if present, is much fainter and narrower than in other extreme starbursts in the early Universe.

\item Our high-resolution observations allowed us to build the first resolved $L_{\rm [CII]} / L_{\rm IR}$ map of a dusty starburst at $z > 4$. Due to the different morphologies of the dust and gas, the $L_{\rm [CII]} / L_{\rm IR}$ ratio is lower in the center of the galaxies and higher in the outer parts, where there is [CII] emission but no continuum is detected at $>3\sigma$. This is consistent with what has been reported in the local Universe. The pixel values on the $L_{\rm [CII]} / L_{\rm IR}$ map suggest that regions with higher star formation have lower associated $L_{\rm [CII]} / L_{\rm IR}$ values.

\item Our results support a scenario where the most distant starbursts were formed in the same highly dissipative mergers of gas-rich galaxies that seem ubiquitous amongst SMGs at $z < 3$, in agreement with recent simulation of IR-bright sources in the early Universe.

\end{itemize}

Systems like SGP38326 are truly important to understand early massive galaxy formation. The SFR of SGP38326 will decline with an $e$-folding timescale of $\tau_{\rm gas} \sim 100 \, {\rm Myr}$. This assumes a Salpeter IMF and $\tau_{\rm gas} = 2 \times M({\rm H_2})/{\rm SFR}$, where the factor of two accounts for the 50\% gas recycling in stellar evolution \citep{Fu2013Natur.498..338F}. This is a simplistic approach since it does not include the possible effect of gas outflow and inflow, but we can assume that they roughly compensate each other during galaxy growth (if there are strong outflows, $\tau_{\rm gas}$ would be even lower). With that timescale, all the gas would have been consumed in $\sim 0.5\,{\rm Gyr}$ (or $\sim 1$ Gyr if Chabrier IMF was assumed). Therefore, our extremely bright interacting starbursts are the likely progenitor of a massive elliptical at $z \sim 3$. We are, consequently, witnessing the formation of the red sequence of galaxies \citep{Kriek2008ApJ...682..896K}.

\begin{acknowledgements}
IO, RJI, ZYZ, LD, AL and SM acknowledge support from the European Research Council in the form of the Advanced Investigator Programme, 321302, {\sc cosmicism}. IRS acknowledges support from STFC (ST/L00075X/1), the ERC Advanced Investigator programme DUSTYGAL 321334 and a Royal Society/Wolfson Merit Award. This paper makes use of the following ALMA data: ADS/JAO.ALMA\#2013.1.00001.S and ADS/JAO.ALMA\#2013.1.00449.S. ALMA is a partnership of ESO (representing its member states), NSF (USA) and NINS (Japan), together with NRC (Canada) and NSC and ASIAA (Taiwan) and KASI (Republic of Korea), in cooperation with the Republic of Chile. The Joint ALMA Observatory is operated by ESO, AUI/NRAO and NAOJ." The {\it H}-ATLAS is a project with {\it Herschel}, which is an ESA space observatory with science instruments provided by European-led Principal Investigator consortia and with important participation from NASA. The {\it H}-ATLAS website is http://www.h-atlas.org/.
\end{acknowledgements}

\bibliographystyle{mn2e}

\bibliography{cii_references}

\begin{thebibliography}{87}
\expandafter\ifx\csname natexlab\endcsname\relax\def\natexlab#1{#1}\fi

\bibitem[{{ALMA Partnership} {et~al}\mbox{.}(2015){ALMA Partnership},
  {Vlahakis}, {Hunter}, {Hodge}, {P{\'e}rez}, {Andreani}, {Brogan}, {Cox},
  {Martin}, {Zwaan}, {Matsushita}, {Dent}, {Impellizzeri}, {Fomalont}, {Asaki},
  {Barkats}, {Hills}, {Hirota}, {Kneissl}, {Liuzzo}, {Lucas}, {Marcelino},
  {Nakanishi}, {Phillips}, {Richards}, {Toledo}, {Aladro}, {Broguiere},
  {Cortes}, {Cortes}, {Espada}, {Galarza}, {Garcia-Appadoo}, {Guzman-Ramirez},
  {Hales}, {Humphreys}, {Jung}, {Kameno}, {Laing}, {Leon}, {Marconi},
  {Mignano}, {Nikolic}, {Nyman}, {Radiszcz}, {Remijan}, {Rod{\'o}n}, {Sawada},
  {Takahashi}, {Tilanus}, {Vila Vilaro}, {Watson}, {Wiklind}, {Ao}, {Di
  Francesco}, {Hatsukade}, {Hatziminaoglou}, {Mangum}, {Matsuda}, {van Kampen},
  {Wootten}, {de Gregorio-Monsalvo}, {Dumas}, {Francke}, {Gallardo}, {Garcia},
  {Gonzalez}, {Hill}, {Iono}, {Kaminski}, {Karim}, {Krips}, {Kurono},
  {Lonsdale}, {Lopez}, {Morales}, {Plarre}, {Videla}, {Villard}, {Hibbard}, \&
  {Tatematsu}}]{Vlahakis2015ApJ...808L...4A}
{ALMA Partnership} {et~al.}, 2015, \apjl, 808, L4

\bibitem[{{Andrews} \& {Thompson}(2011)}]{Andrews2011ApJ...727...97A}
{Andrews} B.~H., {Thompson} T.~A., 2011, \apj, 727, 97

\bibitem[{{Appleton} {et~al}\mbox{.}(2013){Appleton}, {Guillard}, {Boulanger},
  {Cluver}, {Ogle}, {Falgarone}, {Pineau des For{\^e}ts}, {O'Sullivan}, {Duc},
  {Gallagher}, {Gao}, {Jarrett}, {Konstantopoulos}, {Lisenfeld}, {Lord}, {Lu},
  {Peterson}, {Struck}, {Sturm}, {Tuffs}, {Valchanov}, {van der Werf}, \&
  {Xu}}]{Appleton2013ApJ...777...66A}
{Appleton} P.~N. {et~al.}, 2013, \apj, 777, 66

\bibitem[{{Appleton} {et~al}\mbox{.}(2015){Appleton}, {Lanz}, {Bitsakis},
  {Wang}, {Peterson}, {Lisenfeld}, {Alatalo}, {Guillard}, {Boulanger},
  {Cluver}, {Gao}, {Helou}, {Ogle}, \& {Struck}}]{Appleton2015ApJ...812..118A}
{Appleton} P.~N. {et~al.}, 2015, \apj, 812, 118

\bibitem[{{Berta} {et~al}\mbox{.}(2015){Berta}, {Lutz}, {Genzel},
  {Foerster-Schreiber}, \& {Tacconi}}]{Berta2015arXiv151105147B}
{Berta} S., {Lutz} D., {Genzel} R., {Foerster-Schreiber} N.~M., {Tacconi}
  L.~J., 2015, ArXiv e-prints

\bibitem[{{Blain} {et~al}\mbox{.}(2002){Blain}, {Smail}, {Ivison}, {Kneib}, \&
  {Frayer}}]{Blain2002PhR...369..111B}
{Blain} A.~W., {Smail} I., {Ivison} R.~J., {Kneib} J.-P., {Frayer} D.~T., 2002,
  \physrep, 369, 111

\bibitem[{{Bothwell} {et~al}\mbox{.}(2010){Bothwell}, {Chapman}, {Tacconi},
  {Smail}, {Ivison}, {Casey}, {Bertoldi}, {Beswick}, {Biggs}, {Blain}, {Cox},
  {Genzel}, {Greve}, {Kennicutt}, {Muxlow}, {Neri}, \&
  {Omont}}]{Bothwell2010MNRAS.405..219B}
{Bothwell} M.~S. {et~al.}, 2010, \mnras, 405, 219

\bibitem[{{Bussmann} {et~al}\mbox{.}(2013){Bussmann}, {P{\'e}rez-Fournon},
  {Amber}, {Calanog}, {Gurwell}, {Dannerbauer}, {De Bernardis}, {Fu}, {Harris},
  {Krips}, {Lapi}, {Maiolino}, {Omont}, {Riechers}, {Wardlow}, {Baker},
  {Birkinshaw}, {Bock}, {Bourne}, {Clements}, {Cooray}, {De Zotti}, {Dunne},
  {Dye}, {Eales}, {Farrah}, {Gavazzi}, {Gonz{\'a}lez Nuevo}, {Hopwood}, {Ibar},
  {Ivison}, {Laporte}, {Maddox}, {Mart{\'{\i}}nez-Navajas}, {Michalowski},
  {Negrello}, {Oliver}, {Roseboom}, {Scott}, {Serjeant}, {Smith}, {Smith},
  {Streblyanska}, {Valiante}, {van der Werf}, {Verma}, {Vieira}, {Wang}, \&
  {Wilner}}]{Bussmann2013ApJ...779...25B}
{Bussmann} R.~S. {et~al.}, 2013, \apj, 779, 25

\bibitem[{{Bussmann} {et~al}\mbox{.}(2015){Bussmann}, {Riechers}, {Fialkov},
  {Scudder}, {Hayward}, {Cowley}, {Bock}, {Calanog}, {Chapman}, {Cooray}, {De
  Bernardis}, {Farrah}, {Fu}, {Gavazzi}, {Hopwood}, {Ivison}, {Jarvis},
  {Lacey}, {Loeb}, {Oliver}, {P{\'e}rez-Fournon}, {Rigopoulou}, {Roseboom},
  {Scott}, {Smith}, {Vieira}, {Wang}, \&
  {Wardlow}}]{Bussmann2015ApJ...812...43B}
{Bussmann} R.~S. {et~al.}, 2015, \apj, 812, 43

\bibitem[{{Carilli} {et~al}\mbox{.}(2013){Carilli}, {Riechers}, {Walter},
  {Maiolino}, {Wagg}, {Lentati}, {McMahon}, \&
  {Wolfe}}]{Carilli2013ApJ...763..120C}
{Carilli} C.~L., {Riechers} D., {Walter} F., {Maiolino} R., {Wagg} J.,
  {Lentati} L., {McMahon} R., {Wolfe} A., 2013, \apj, 763, 120

\bibitem[{{Carilli} \& {Walter}(2013)}]{Carilli2013ARA&A..51..105C}
{Carilli} C.~L., {Walter} F., 2013, \araa, 51, 105

\bibitem[{{Casey} {et~al}\mbox{.}(2013){Casey}, {Chen}, {Cowie}, {Barger},
  {Capak}, {Ilbert}, {Koss}, {Lee}, {Le Floc'h}, {Sanders}, \&
  {Williams}}]{Casey2013MNRAS.436.1919C}
{Casey} C.~M. {et~al.}, 2013, \mnras, 436, 1919

\bibitem[{{Chapman} {et~al}\mbox{.}(2005){Chapman}, {Blain}, {Smail}, \&
  {Ivison}}]{Chapman2005ApJ...622..772C}
{Chapman} S.~C., {Blain} A.~W., {Smail} I., {Ivison} R.~J., 2005, \apj, 622,
  772

\bibitem[{{Cicone} {et~al}\mbox{.}(2015){Cicone}, {Maiolino}, {Gallerani},
  {Neri}, {Ferrara}, {Sturm}, {Fiore}, {Piconcelli}, \&
  {Feruglio}}]{Cicone2015A&A...574A..14C}
{Cicone} C. {et~al.}, 2015, \aap, 574, A14

\bibitem[{{Cooray} {et~al}\mbox{.}(2014){Cooray}, {Calanog}, {Wardlow}, {Bock},
  {Bridge}, {Burgarella}, {Bussmann}, {Casey}, {Clements}, {Conley}, {Farrah},
  {Fu}, {Gavazzi}, {Ivison}, {La Porte}, {Lo Faro}, {Ma}, {Magdis}, {Oliver},
  {Osage}, {P{\'e}rez-Fournon}, {Riechers}, {Rigopoulou}, {Scott}, {Viero}, \&
  {Watson}}]{Cooray2014ApJ...790...40C}
{Cooray} A. {et~al.}, 2014, \apj, 790, 40

\bibitem[{{Dame}(2011)}]{Dame2011arXiv1101.1499D}
{Dame} T.~M., 2011, ArXiv e-prints

\bibitem[{{De Breuck} {et~al}\mbox{.}(2014){De Breuck}, {Williams}, {Swinbank},
  {Caselli}, {Coppin}, {Davis}, {Maiolino}, {Nagao}, {Smail}, {Walter},
  {Wei{\ss}}, \& {Zwaan}}]{DeBreuck2014A&A...565A..59D}
{De Breuck} C. {et~al.}, 2014, \aap, 565, A59

\bibitem[{{D{\'{\i}}az-Santos} {et~al}\mbox{.}(2014){D{\'{\i}}az-Santos},
  {Armus}, {Charmandaris}, {Stacey}, {Murphy}, {Haan}, {Stierwalt}, {Malhotra},
  {Appleton}, {Inami}, {Magdis}, {Elbaz}, {Evans}, {Mazzarella}, {Surace}, {van
  der Werf}, {Xu}, {Lu}, {Meijerink}, {Howell}, {Petric}, {Veilleux}, \&
  {Sanders}}]{Tanio2014ApJ...788L..17D}
{D{\'{\i}}az-Santos} T. {et~al.}, 2014, \apjl, 788, L17

\bibitem[{{D{\'{\i}}az-Santos} {et~al}\mbox{.}(2013){D{\'{\i}}az-Santos},
  {Armus}, {Charmandaris}, {Stierwalt}, {Murphy}, {Haan}, {Inami}, {Malhotra},
  {Meijerink}, {Stacey}, {Petric}, {Evans}, {Veilleux}, {van der Werf}, {Lord},
  {Lu}, {Howell}, {Appleton}, {Mazzarella}, {Surace}, {Xu}, {Schulz},
  {Sanders}, {Bridge}, {Chan}, {Frayer}, {Iwasawa}, {Melbourne}, \&
  {Sturm}}]{DiazSantos2013ApJ...774...68D}
{D{\'{\i}}az-Santos} T. {et~al.}, 2013, \apj, 774, 68

\bibitem[{{Diaz-Santos} {et~al}\mbox{.}(2015){Diaz-Santos}, {Assef}, {Blain},
  {Tsai}, {Aravena}, {Eisenhardt}, {Wu}, {Stern}, \&
  {Bridge}}]{Tanio2015arXiv151104079D}
{Diaz-Santos} T. {et~al.}, 2015, ArXiv e-prints

\bibitem[{{Downes} \& {Solomon}(1998)}]{Downes1998ApJ...507..615D}
{Downes} D., {Solomon} P.~M., 1998, \apj, 507, 615

\bibitem[{{Dunne} \& {Eales}(2001)}]{Dunne2001MNRAS.327..697D}
{Dunne} L., {Eales} S.~A., 2001, \mnras, 327, 697

\bibitem[{{Dye} {et~al}\mbox{.}(2015){Dye}, {Furlanetto}, {Swinbank},
  {Vlahakis}, {Nightingale}, {Dunne}, {Eales}, {Smail}, {Oteo}, {Hunter},
  {Negrello}, {Dannerbauer}, {Ivison}, {Gavazzi}, {Cooray}, \& {van der
  Werf}}]{Dye2015MNRAS.452.2258D}
{Dye} S. {et~al.}, 2015, \mnras, 452, 2258

\bibitem[{{Eales} {et~al}\mbox{.}(2010){Eales}, {Dunne}, {Clements}, {Cooray},
  {de Zotti}, {Dye}, {Ivison}, {Jarvis}, {Lagache}, {Maddox}, {Negrello},
  {Serjeant}, {Thompson}, {van Kampen}, {Amblard}, {Andreani}, {Baes},
  {Beelen}, {Bendo}, {Benford}, {Bertoldi}, {Bock}, {Bonfield}, {Boselli},
  {Bridge}, {Buat}, {Burgarella}, {Carlberg}, {Cava}, {Chanial}, {Charlot},
  {Christopher}, {Coles}, {Cortese}, {Dariush}, {da Cunha}, {Dalton}, {Danese},
  {Dannerbauer}, {Driver}, {Dunlop}, {Fan}, {Farrah}, {Frayer}, {Frenk},
  {Geach}, {Gardner}, {Gomez}, {Gonz{\'a}lez-Nuevo}, {Gonz{\'a}lez-Solares},
  {Griffin}, {Hardcastle}, {Hatziminaoglou}, {Herranz}, {Hughes}, {Ibar},
  {Jeong}, {Lacey}, {Lapi}, {Lawrence}, {Lee}, {Leeuw}, {Liske},
  {L{\'o}pez-Caniego}, {M{\"u}ller}, {Nandra}, {Panuzzo}, {Papageorgiou},
  {Patanchon}, {Peacock}, {Pearson}, {Phillipps}, {Pohlen}, {Popescu},
  {Rawlings}, {Rigby}, {Rigopoulou}, {Robotham}, {Rodighiero}, {Sansom},
  {Schulz}, {Scott}, {Smith}, {Sibthorpe}, {Smail}, {Stevens}, {Sutherland},
  {Takeuchi}, {Tedds}, {Temi}, {Tuffs}, {Trichas}, {Vaccari}, {Valtchanov},
  {van der Werf}, {Verma}, {Vieria}, {Vlahakis}, \&
  {White}}]{Eales2010PASP..122..499E}
{Eales} S. {et~al.}, 2010, \pasp, 122, 499

\bibitem[{{Edge} {et~al}\mbox{.}(2013){Edge}, {Sutherland}, {Kuijken},
  {Driver}, {McMahon}, {Eales}, \& {Emerson}}]{Edge2013Msngr.154...32E}
{Edge} A., {Sutherland} W., {Kuijken} K., {Driver} S., {McMahon} R., {Eales}
  S., {Emerson} J.~P., 2013, The Messenger, 154, 32

\bibitem[{{Elbaz} {et~al}\mbox{.}(2011){Elbaz}, {Dickinson}, {Hwang},
  {D{\'{\i}}az-Santos}, {Magdis}, {Magnelli}, {Le Borgne}, {Galliano},
  {Pannella}, {Chanial}, {Armus}, {Charmandaris}, {Daddi}, {Aussel}, {Popesso},
  {Kartaltepe}, {Altieri}, {Valtchanov}, {Coia}, {Dannerbauer}, {Dasyra},
  {Leiton}, {Mazzarella}, {Alexander}, {Buat}, {Burgarella}, {Chary}, {Gilli},
  {Ivison}, {Juneau}, {Le Floc'h}, {Lutz}, {Morrison}, {Mullaney}, {Murphy},
  {Pope}, {Scott}, {Brodwin}, {Calzetti}, {Cesarsky}, {Charlot}, {Dole},
  {Eisenhardt}, {Ferguson}, {F{\"o}rster Schreiber}, {Frayer}, {Giavalisco},
  {Huynh}, {Koekemoer}, {Papovich}, {Reddy}, {Surace}, {Teplitz}, {Yun}, \&
  {Wilson}}]{Elbaz2011A&A...533A.119E}
{Elbaz} D. {et~al.}, 2011, \aap, 533, A119

\bibitem[{{Engel} {et~al}\mbox{.}(2010){Engel}, {Tacconi}, {Davies}, {Neri},
  {Smail}, {Chapman}, {Genzel}, {Cox}, {Greve}, {Ivison}, {Blain}, {Bertoldi},
  \& {Omont}}]{Engel2010ApJ...724..233E}
{Engel} H. {et~al.}, 2010, \apj, 724, 233

\bibitem[{{Farrah} {et~al}\mbox{.}(2013){Farrah}, {Lebouteiller}, {Spoon},
  {Bernard-Salas}, {Pearson}, {Rigopoulou}, {Smith}, {Gonz{\'a}lez-Alfonso},
  {Clements}, {Efstathiou}, {Cormier}, {Afonso}, {Petty}, {Harris}, {Hurley},
  {Borys}, {Verma}, {Cooray}, \& {Salvatelli}}]{Farrah2013ApJ...776...38F}
{Farrah} D. {et~al.}, 2013, \apj, 776, 38

\bibitem[{{F{\"o}rster Schreiber} {et~al}\mbox{.}(2009){F{\"o}rster Schreiber},
  {Genzel}, {Bouch{\'e}}, {Cresci}, {Davies}, {Buschkamp}, {Shapiro},
  {Tacconi}, {Hicks}, {Genel}, {Shapley}, {Erb}, {Steidel}, {Lutz},
  {Eisenhauer}, {Gillessen}, {Sternberg}, {Renzini}, {Cimatti}, {Daddi},
  {Kurk}, {Lilly}, {Kong}, {Lehnert}, {Nesvadba}, {Verma}, {McCracken},
  {Arimoto}, {Mignoli}, \& {Onodera}}]{Forster2009ApJ...706.1364F}
{F{\"o}rster Schreiber} N.~M. {et~al.}, 2009, \apj, 706, 1364

\bibitem[{{Fu} {et~al}\mbox{.}(2013){Fu}, {Cooray}, {Feruglio}, {Ivison},
  {Riechers}, {Gurwell}, {Bussmann}, {Harris}, {Altieri}, {Aussel}, {Baker},
  {Bock}, {Boylan-Kolchin}, {Bridge}, {Calanog}, {Casey}, {Cava}, {Chapman},
  {Clements}, {Conley}, {Cox}, {Farrah}, {Frayer}, {Hopwood}, {Jia}, {Magdis},
  {Marsden}, {Mart{\'{\i}}nez-Navajas}, {Negrello}, {Neri}, {Oliver}, {Omont},
  {Page}, {P{\'e}rez-Fournon}, {Schulz}, {Scott}, {Smith}, {Vaccari},
  {Valtchanov}, {Vieira}, {Viero}, {Wang}, {Wardlow}, \&
  {Zemcov}}]{Fu2013Natur.498..338F}
{Fu} H. {et~al.}, 2013, \nat, 498, 338

\bibitem[{{Gabor} \& {Dav{\'e}}(2012)}]{Gabor2012MNRAS.427.1816G}
{Gabor} J.~M., {Dav{\'e}} R., 2012, \mnras, 427, 1816

\bibitem[{{Geach} {et~al}\mbox{.}(2013){Geach}, {Chapin}, {Coppin}, {Dunlop},
  {Halpern}, {Smail}, {van der Werf}, {Serjeant}, {Farrah}, {Roseboom},
  {Targett}, {Arumugam}, {Asboth}, {Blain}, {Chrysostomou}, {Clarke}, {Ivison},
  {Jones}, {Karim}, {Mackenzie}, {Meijerink}, {Micha{\l}owski}, {Scott},
  {Simpson}, {Swinbank}, {Alexander}, {Almaini}, {Aretxaga}, {Best}, {Chapman},
  {Clements}, {Conselice}, {Danielson}, {Eales}, {Edge}, {Gibb}, {Hughes},
  {Jenness}, {Knudsen}, {Lacey}, {Marsden}, {McMahon}, {Oliver}, {Page},
  {Peacock}, {Rigopoulou}, {Robson}, {Spaans}, {Stevens}, {Webb}, {Willott},
  {Wilson}, \& {Zemcov}}]{Geach2013MNRAS.432...53G}
{Geach} J.~E. {et~al.}, 2013, \mnras, 432, 53

\bibitem[{{Genzel} {et~al}\mbox{.}(2014){Genzel}, {F{\"o}rster Schreiber},
  {Lang}, {Tacchella}, {Tacconi}, {Wuyts}, {Bandara}, {Burkert}, {Buschkamp},
  {Carollo}, {Cresci}, {Davies}, {Eisenhauer}, {Hicks}, {Kurk}, {Lilly},
  {Lutz}, {Mancini}, {Naab}, {Newman}, {Peng}, {Renzini}, {Shapiro Griffin},
  {Sternberg}, {Vergani}, {Wisnioski}, {Wuyts}, \&
  {Zamorani}}]{Genzel2014ApJ...785...75G}
{Genzel} R. {et~al.}, 2014, \apj, 785, 75

\bibitem[{{Hartley} {et~al}\mbox{.}(2013){Hartley}, {Almaini}, {Mortlock},
  {Conselice}, {Gr{\"u}tzbauch}, {Simpson}, {Bradshaw}, {Chuter}, {Foucaud},
  {Cirasuolo}, {Dunlop}, {McLure}, \& {Pearce}}]{Hartley2013MNRAS.431.3045H}
{Hartley} W.~G. {et~al.}, 2013, \mnras, 431, 3045

\bibitem[{{Hodge} {et~al}\mbox{.}(2012){Hodge}, {Carilli}, {Walter}, {de Blok},
  {Riechers}, {Daddi}, \& {Lentati}}]{Hodge2012ApJ...760...11H}
{Hodge} J.~A., {Carilli} C.~L., {Walter} F., {de Blok} W.~J.~G., {Riechers} D.,
  {Daddi} E., {Lentati} L., 2012, \apj, 760, 11

\bibitem[{{Hodge} {et~al}\mbox{.}(2013){Hodge}, {Karim}, {Smail}, {Swinbank},
  {Walter}, {Biggs}, {Ivison}, {Weiss}, {Alexander}, {Bertoldi}, {Brandt},
  {Chapman}, {Coppin}, {Cox}, {Danielson}, {Dannerbauer}, {De Breuck},
  {Decarli}, {Edge}, {Greve}, {Knudsen}, {Menten}, {Rix}, {Schinnerer},
  {Simpson}, {Wardlow}, \& {van der Werf}}]{Hodge2013ApJ...768...91H}
{Hodge} J.~A. {et~al.}, 2013, \apj, 768, 91

\bibitem[{{Hodge} {et~al}\mbox{.}(2015){Hodge}, {Riechers}, {Decarli},
  {Walter}, {Carilli}, {Daddi}, \& {Dannerbauer}}]{Hodge2015ApJ...798L..18H}
{Hodge} J.~A., {Riechers} D., {Decarli} R., {Walter} F., {Carilli} C.~L.,
  {Daddi} E., {Dannerbauer} H., 2015, \apjl, 798, L18

\bibitem[{{Holland} {et~al}\mbox{.}(2013){Holland}, {Bintley}, {Chapin},
  {Chrysostomou}, {Davis}, {Dempsey}, {Duncan}, {Fich}, {Friberg}, {Halpern},
  {Irwin}, {Jenness}, {Kelly}, {MacIntosh}, {Robson}, {Scott}, {Ade},
  {Atad-Ettedgui}, {Berry}, {Craig}, {Gao}, {Gibb}, {Hilton}, {Hollister},
  {Kycia}, {Lunney}, {McGregor}, {Montgomery}, {Parkes}, {Tilanus}, {Ullom},
  {Walther}, {Walton}, {Woodcraft}, {Amiri}, {Atkinson}, {Burger}, {Chuter},
  {Coulson}, {Doriese}, {Dunare}, {Economou}, {Niemack}, {Parsons},
  {Reintsema}, {Sibthorpe}, {Smail}, {Sudiwala}, \&
  {Thomas}}]{Holland2013MNRAS.430.2513H}
{Holland} W.~S. {et~al.}, 2013, \mnras, 430, 2513

\bibitem[{{Hwang} {et~al}\mbox{.}(2010){Hwang}, {Elbaz}, {Magdis}, {Daddi},
  {Symeonidis}, {Altieri}, {Amblard}, {Andreani}, {Arumugam}, {Auld}, {Aussel},
  {Babbedge}, {Berta}, {Blain}, {Bock}, {Bongiovanni}, {Boselli}, {Buat},
  {Burgarella}, {Castro-Rodr{\'{\i}}guez}, {Cava}, {Cepa}, {Chanial}, {Chapin},
  {Chary}, {Cimatti}, {Clements}, {Conley}, {Conversi}, {Cooray},
  {Dannerbauer}, {Dickinson}, {Dominguez}, {Dowell}, {Dunlop}, {Dwek}, {Eales},
  {Farrah}, {Schreiber}, {Fox}, {Franceschini}, {Gear}, {Genzel}, {Glenn},
  {Griffin}, {Gruppioni}, {Halpern}, {Hatziminaoglou}, {Ibar}, {Isaak},
  {Ivison}, {Jeong}, {Lagache}, {Le Borgne}, {Le Floc'h}, {Lee}, {Lee}, {Lee},
  {Levenson}, {Lu}, {Lutz}, {Madden}, {Maffei}, {Magnelli}, {Mainetti},
  {Maiolino}, {Marchetti}, {Mortier}, {Nguyen}, {Nordon}, {O'Halloran},
  {Okumura}, {Oliver}, {Omont}, {Page}, {Panuzzo}, {Papageorgiou}, {Pearson},
  {P{\'e}rez-Fournon}, {Garc{\'{\i}}a}, {Poglitsch}, {Pohlen}, {Popesso},
  {Pozzi}, {Rawlings}, {Rigopoulou}, {Riguccini}, {Rizzo}, {Rodighiero},
  {Roseboom}, {Rowan-Robinson}, {Saintonge}, {Portal}, {Santini}, {Sauvage},
  {Schulz}, {Scott}, {Seymour}, {Shao}, {Shupe}, {Smith}, {Stevens}, {Sturm},
  {Tacconi}, {Trichas}, {Tugwell}, {Vaccari}, {Valtchanov}, {Vieira},
  {Vigroux}, {Wang}, {Ward}, {Wright}, {Xu}, \&
  {Zemcov}}]{Hwang2010MNRAS.409...75H}
{Hwang} H.~S. {et~al.}, 2010, \mnras, 409, 75

\bibitem[{{Ikarashi} {et~al}\mbox{.}(2015){Ikarashi}, {Ivison}, {Caputi},
  {Aretxaga}, {Dunlop}, {Hatsukade}, {Hughes}, {Iono}, {Izumi}, {Kawabe},
  {Kohno}, {Lagos}, {Motohara}, {Nakanishi}, {Ohta}, {Tamura}, {Umehata},
  {Wilson}, {Yabe}, \& {Yun}}]{Ikarashi2015ApJ...810..133I}
{Ikarashi} S. {et~al.}, 2015, \apj, 810, 133

\bibitem[{{Ivison} {et~al}\mbox{.}(2011){Ivison}, {Papadopoulos}, {Smail},
  {Greve}, {Thomson}, {Xilouris}, \& {Chapman}}]{Ivison2011MNRAS.412.1913I}
{Ivison} R.~J., {Papadopoulos} P.~P., {Smail} I., {Greve} T.~R., {Thomson}
  A.~P., {Xilouris} E.~M., {Chapman} S.~C., 2011, \mnras, 412, 1913

\bibitem[{{Ivison} {et~al}\mbox{.}(2013){Ivison}, {Swinbank}, {Smail},
  {Harris}, {Bussmann}, {Cooray}, {Cox}, {Fu}, {Kov{\'a}cs}, {Krips},
  {Narayanan}, {Negrello}, {Neri}, {Pe{\~n}arrubia}, {Richard}, {Riechers},
  {Rowlands}, {Staguhn}, {Targett}, {Amber}, {Baker}, {Bourne}, {Bertoldi},
  {Bremer}, {Calanog}, {Clements}, {Dannerbauer}, {Dariush}, {De Zotti},
  {Dunne}, {Eales}, {Farrah}, {Fleuren}, {Franceschini}, {Geach}, {George},
  {Helly}, {Hopwood}, {Ibar}, {Jarvis}, {Kneib}, {Maddox}, {Omont}, {Scott},
  {Serjeant}, {Smith}, {Thompson}, {Valiante}, {Valtchanov}, {Vieira}, \& {van
  der Werf}}]{Ivison2013ApJ...772..137I}
{Ivison} R.~J. {et~al.}, 2013, \apj, 772, 137

\bibitem[{{Karim} {et~al}\mbox{.}(2013){Karim}, {Swinbank}, {Hodge}, {Smail},
  {Walter}, {Biggs}, {Simpson}, {Danielson}, {Alexander}, {Bertoldi}, {de
  Breuck}, {Chapman}, {Coppin}, {Dannerbauer}, {Edge}, {Greve}, {Ivison},
  {Knudsen}, {Menten}, {Schinnerer}, {Wardlow}, {Wei{\ss}}, \& {van der
  Werf}}]{Karim2013MNRAS.432....2K}
{Karim} A. {et~al.}, 2013, \mnras, 432, 2

\bibitem[{{Kennicutt}(1998)}]{Kennicutt1998ARA&A..36..189K}
{Kennicutt}, Jr. R.~C., 1998, \araa, 36, 189

\bibitem[{{Kodama} {et~al}\mbox{.}(2007){Kodama}, {Tanaka}, {Kajisawa}, {Kurk},
  {Venemans}, {De Breuck}, {Vernet}, \& {Lidman}}]{Kodama2007MNRAS.377.1717K}
{Kodama} T., {Tanaka} I., {Kajisawa} M., {Kurk} J., {Venemans} B., {De Breuck}
  C., {Vernet} J., {Lidman} C., 2007, \mnras, 377, 1717

\bibitem[{{Kriek} {et~al}\mbox{.}(2008){Kriek}, {van der Wel}, {van Dokkum},
  {Franx}, \& {Illingworth}}]{Kriek2008ApJ...682..896K}
{Kriek} M., {van der Wel} A., {van Dokkum} P.~G., {Franx} M., {Illingworth}
  G.~D., 2008, \apj, 682, 896

\bibitem[{{Leroy} {et~al}\mbox{.}(2011){Leroy}, {Bolatto}, {Gordon},
  {Sandstrom}, {Gratier}, {Rosolowsky}, {Engelbracht}, {Mizuno}, {Corbelli},
  {Fukui}, \& {Kawamura}}]{Leroy2011ApJ...737...12L}
{Leroy} A.~K. {et~al.}, 2011, \apj, 737, 12

\bibitem[{{Lutz} {et~al}\mbox{.}(2015){Lutz}, {Berta}, {Contursi}, {F{\"o}rster
  Schreiber}, {Genzel}, {Graci{\'a}-Carpio}, {Herrera-Camus}, {Netzer},
  {Sturm}, {Tacconi}, {Tadaki}, \& {Veilleux}}]{Lutz2015arXiv151102075L}
{Lutz} D. {et~al.}, 2015, ArXiv e-prints

\bibitem[{{Madden} {et~al}\mbox{.}(1997){Madden}, {Poglitsch}, {Geis},
  {Stacey}, \& {Townes}}]{Madden1997ApJ...483..200M}
{Madden} S.~C., {Poglitsch} A., {Geis} N., {Stacey} G.~J., {Townes} C.~H.,
  1997, \apj, 483, 200

\bibitem[{{Magdis} {et~al}\mbox{.}(2012){Magdis}, {Daddi}, {B{\'e}thermin},
  {Sargent}, {Elbaz}, {Pannella}, {Dickinson}, {Dannerbauer}, {da Cunha},
  {Walter}, {Rigopoulou}, {Charmandaris}, {Hwang}, \&
  {Kartaltepe}}]{Magdis2012ApJ...760....6M}
{Magdis} G.~E. {et~al.}, 2012, \apj, 760, 6

\bibitem[{{Magdis} {et~al}\mbox{.}(2011){Magdis}, {Daddi}, {Elbaz}, {Sargent},
  {Dickinson}, {Dannerbauer}, {Aussel}, {Walter}, {Hwang}, {Charmandaris},
  {Hodge}, {Riechers}, {Rigopoulou}, {Carilli}, {Pannella}, {Mullaney},
  {Leiton}, \& {Scott}}]{Magdis2011ApJ...740L..15M}
{Magdis} G.~E. {et~al.}, 2011, \apjl, 740, L15

\bibitem[{{Mei} {et~al}\mbox{.}(2009){Mei}, {Holden}, {Blakeslee}, {Ford},
  {Franx}, {Homeier}, {Illingworth}, {Jee}, {Overzier}, {Postman}, {Rosati},
  {Van der Wel}, \& {Bartlett}}]{Mei2009ApJ...690...42M}
{Mei} S. {et~al.}, 2009, \apj, 690, 42

\bibitem[{{Narayanan} {et~al}\mbox{.}(2015){Narayanan}, {Turk}, {Feldmann},
  {Robitaille}, {Hopkins}, {Thompson}, {Hayward}, {Ball},
  {Faucher-Gigu{\`e}re}, \& {Kere{\v s}}}]{Narayanan2015Natur.525..496N}
{Narayanan} D. {et~al.}, 2015, \nat, 525, 496

\bibitem[{{Oke} \& {Gunn}(1983)}]{Oke1983}
{Oke} J.~B., {Gunn} J.~E., 1983, \apj, 266, 713

\bibitem[{{Oliver} {et~al}\mbox{.}(2010){Oliver}, {Wang}, {Smith}, {Altieri},
  {Amblard}, {Arumugam}, {Auld}, {Aussel}, {Babbedge}, {Blain}, {Bock},
  {Boselli}, {Buat}, {Burgarella}, {Castro-Rodr{\'{\i}}guez}, {Cava},
  {Chanial}, {Clements}, {Conley}, {Conversi}, {Cooray}, {Dowell}, {Dwek},
  {Eales}, {Elbaz}, {Fox}, {Franceschini}, {Gear}, {Glenn}, {Griffin},
  {Halpern}, {Hatziminaoglou}, {Ibar}, {Isaak}, {Ivison}, {Lagache},
  {Levenson}, {Lu}, {Madden}, {Maffei}, {Mainetti}, {Marchetti},
  {Mitchell-Wynne}, {Mortier}, {Nguyen}, {O'Halloran}, {Omont}, {Page},
  {Panuzzo}, {Papageorgiou}, {Pearson}, {P{\'e}rez-Fournon}, {Pohlen},
  {Rawlings}, {Raymond}, {Rigopoulou}, {Rizzo}, {Roseboom}, {Rowan-Robinson},
  {S{\'a}nchez Portal}, {Savage}, {Schulz}, {Scott}, {Seymour}, {Shupe},
  {Stevens}, {Symeonidis}, {Trichas}, {Tugwell}, {Vaccari}, {Valiante},
  {Valtchanov}, {Vieira}, {Vigroux}, {Ward}, {Wright}, {Xu}, \&
  {Zemcov}}]{Oliver2010A&A...518L..21O}
{Oliver} S.~J. {et~al.}, 2010, \aap, 518, L21

\bibitem[{{Oteo} {et~al}\mbox{.}(2015){Oteo}, {Zwaan}, {Ivison}, {Smail}, \&
  {Biggs}}]{Oteo2015arXiv150805099O}
{Oteo} I., {Zwaan} M.~A., {Ivison} R.~J., {Smail} I., {Biggs} A.~D., 2015,
  ArXiv e-prints

\bibitem[{{Rawle} {et~al}\mbox{.}(2014){Rawle}, {Egami}, {Bussmann}, {Gurwell},
  {Ivison}, {Boone}, {Combes}, {Danielson}, {Rex}, {Richard}, {Smail},
  {Swinbank}, {Altieri}, {Blain}, {Clement}, {Dessauges-Zavadsky}, {Edge},
  {Fazio}, {Jones}, {Kneib}, {Omont}, {P{\'e}rez-Gonz{\'a}lez}, {Schaerer},
  {Valtchanov}, {van der Werf}, {Walth}, {Zamojski}, \&
  {Zemcov}}]{Rawle2014ApJ...783...59R}
{Rawle} T.~D. {et~al.}, 2014, \apj, 783, 59

\bibitem[{{Riechers} {et~al}\mbox{.}(2013){Riechers}, {Bradford}, {Clements},
  {Dowell}, {P{\'e}rez-Fournon}, {Ivison}, {Bridge}, {Conley}, {Fu}, {Vieira},
  {Wardlow}, {Calanog}, {Cooray}, {Hurley}, {Neri}, {Kamenetzky}, {Aguirre},
  {Altieri}, {Arumugam}, {Benford}, {B{\'e}thermin}, {Bock}, {Burgarella},
  {Cabrera-Lavers}, {Chapman}, {Cox}, {Dunlop}, {Earle}, {Farrah}, {Ferrero},
  {Franceschini}, {Gavazzi}, {Glenn}, {Solares}, {Gurwell}, {Halpern},
  {Hatziminaoglou}, {Hyde}, {Ibar}, {Kov{\'a}cs}, {Krips}, {Lupu}, {Maloney},
  {Martinez-Navajas}, {Matsuhara}, {Murphy}, {Naylor}, {Nguyen}, {Oliver},
  {Omont}, {Page}, {Petitpas}, {Rangwala}, {Roseboom}, {Scott}, {Smith},
  {Staguhn}, {Streblyanska}, {Thomson}, {Valtchanov}, {Viero}, {Wang},
  {Zemcov}, \& {Zmuidzinas}}]{Riechers2013Natur.496..329R}
{Riechers} D.~A. {et~al.}, 2013, \nat, 496, 329

\bibitem[{{Riechers} {et~al}\mbox{.}(2014){Riechers}, {Carilli}, {Capak},
  {Scoville}, {Smol{\v c}i{\'c}}, {Schinnerer}, {Yun}, {Cox}, {Bertoldi},
  {Karim}, \& {Yan}}]{Riechers2014ApJ...796...84R}
{Riechers} D.~A. {et~al.}, 2014, \apj, 796, 84

\bibitem[{{Rosati} {et~al}\mbox{.}(2009){Rosati}, {Tozzi}, {Gobat}, {Santos},
  {Nonino}, {Demarco}, {Lidman}, {Mullis}, {Strazzullo}, {B{\"o}hringer},
  {Fassbender}, {Dawson}, {Tanaka}, {Jee}, {Ford}, {Lamer}, \&
  {Schwope}}]{Rosati2009A&A...508..583R}
{Rosati} P. {et~al.}, 2009, \aap, 508, 583

\bibitem[{{Rowlands} {et~al}\mbox{.}(2014){Rowlands}, {Dunne}, {Dye},
  {Arag{\'o}n-Salamanca}, {Maddox}, {da Cunha}, {Smith}, {Bourne}, {Eales},
  {Gomez}, {Smail}, {Alpaslan}, {Clark}, {Driver}, {Ibar}, {Ivison},
  {Robotham}, {Smith}, \& {Valiante}}]{Rowlands2014MNRAS.441.1017R}
{Rowlands} K. {et~al.}, 2014, \mnras, 441, 1017

\bibitem[{{Sandstrom} {et~al}\mbox{.}(2013){Sandstrom}, {Leroy}, {Walter},
  {Bolatto}, {Croxall}, {Draine}, {Wilson}, {Wolfire}, {Calzetti}, {Kennicutt},
  {Aniano}, {Donovan Meyer}, {Usero}, {Bigiel}, {Brinks}, {de Blok}, {Crocker},
  {Dale}, {Engelbracht}, {Galametz}, {Groves}, {Hunt}, {Koda}, {Kreckel},
  {Linz}, {Meidt}, {Pellegrini}, {Rix}, {Roussel}, {Schinnerer}, {Schruba},
  {Schuster}, {Skibba}, {van der Laan}, {Appleton}, {Armus}, {Brandl},
  {Gordon}, {Hinz}, {Krause}, {Montiel}, {Sauvage}, {Schmiedeke}, {Smith}, \&
  {Vigroux}}]{Sandstrom2013ApJ...777....5S}
{Sandstrom} K.~M. {et~al.}, 2013, \apj, 777, 5

\bibitem[{{Scoville} {et~al}\mbox{.}(2015){Scoville}, {Sheth}, {Aussel},
  {Vanden Bout}, {Capak}, {Bongiorno}, {Casey}, {Murchikova}, {Koda},
  {'Alvarez-M'arquez}, {Lee}, {Laigle}, {McCracken}, {Ilbert}, {Pope},
  {Sanders}, {Chu}, {Toft}, {Ivison}, \&
  {Manohar}}]{Scoville2015arXiv151105149S}
{Scoville} N. {et~al.}, 2015, ArXiv e-prints

\bibitem[{{Simpson} {et~al}\mbox{.}(2015{\natexlab{a}}){Simpson}, {Smail},
  {Swinbank}, {Almaini}, {Blain}, {Bremer}, {Chapman}, {Chen}, {Conselice},
  {Coppin}, {Danielson}, {Dunlop}, {Edge}, {Farrah}, {Geach}, {Hartley},
  {Ivison}, {Karim}, {Lani}, {Ma}, {Meijerink}, {Micha{\l}owski}, {Mortlock},
  {Scott}, {Simpson}, {Spaans}, {Thomson}, {van Kampen}, \& {van der
  Werf}}]{Simpson2015ApJ...799...81S}
{Simpson} J.~M. {et~al.}, 2015{\natexlab{a}}, \apj, 799, 81

\bibitem[{{Simpson} {et~al}\mbox{.}(2015{\natexlab{b}}){Simpson}, {Smail},
  {Swinbank}, {Chapman}, {Geach}, {Ivison}, {Thomson}, {Aretxaga}, {Blain},
  {Cowley}, {Chen}, {Coppin}, {Dunlop}, {Edge}, {Farrah}, {Ibar}, {Karim},
  {Knudsen}, {Meijerink}, {Micha{\l}owski}, {Scott}, {Spaans}, \& {van der
  Werf}}]{Simpson2015ApJ...807..128S}
{Simpson} J.~M. {et~al.}, 2015{\natexlab{b}}, \apj, 807, 128

\bibitem[{{Siringo} {et~al}\mbox{.}(2009){Siringo}, {Kreysa}, {Kov{\'a}cs},
  {Schuller}, {Wei{\ss}}, {Esch}, {Gem{\"u}nd}, {Jethava}, {Lundershausen},
  {Colin}, {G{\"u}sten}, {Menten}, {Beelen}, {Bertoldi}, {Beeman}, \&
  {Haller}}]{Siringo2009A&A...497..945S}
{Siringo} G. {et~al.}, 2009, \aap, 497, 945

\bibitem[{{Sobral} {et~al}\mbox{.}(2013){Sobral}, {Smail}, {Best}, {Geach},
  {Matsuda}, {Stott}, {Cirasuolo}, \& {Kurk}}]{Sobral2013MNRAS.428.1128S}
{Sobral} D., {Smail} I., {Best} P.~N., {Geach} J.~E., {Matsuda} Y., {Stott}
  J.~P., {Cirasuolo} M., {Kurk} J., 2013, \mnras, 428, 1128

\bibitem[{{Solomon} {et~al}\mbox{.}(1997){Solomon}, {Downes}, {Radford}, \&
  {Barrett}}]{Solomon1997ApJ...478..144S}
{Solomon} P.~M., {Downes} D., {Radford} S.~J.~E., {Barrett} J.~W., 1997, \apj,
  478, 144

\bibitem[{{Stacey} {et~al}\mbox{.}(1999){Stacey}, {Swain}, {Bradford},
  {Barlow}, {Cox}, {Fischer}, {Lord}, {Luhman}, {Rieu}, {Smith}, {Unger}, \&
  {Wolfire}}]{Stacey1999ESASP.427..973S}
{Stacey} G.~J. {et~al.}, 1999, in ESA Special Publication, Vol. 427, The
  Universe as Seen by ISO, {Cox} P., {Kessler} M., eds., p. 973

\bibitem[{{Stanford} {et~al}\mbox{.}(2006){Stanford}, {Romer}, {Sabirli},
  {Davidson}, {Hilton}, {Viana}, {Collins}, {Kay}, {Liddle}, {Mann}, {Miller},
  {Nichol}, {West}, {Conselice}, {Spinrad}, {Stern}, \&
  {Bundy}}]{Stanford2006ApJ...646L..13S}
{Stanford} S.~A. {et~al.}, 2006, \apjl, 646, L13

\bibitem[{{Strazzullo} {et~al}\mbox{.}(2010){Strazzullo}, {Rosati}, {Pannella},
  {Gobat}, {Santos}, {Nonino}, {Demarco}, {Lidman}, {Tanaka}, {Mullis},
  {Nu{\~n}ez}, {Rettura}, {Jee}, {B{\"o}hringer}, {Bender}, {Bouwens},
  {Dawson}, {Fassbender}, {Franx}, {Perlmutter}, \&
  {Postman}}]{Strazzullo2010A&A...524A..17S}
{Strazzullo} V. {et~al.}, 2010, \aap, 524, A17

\bibitem[{{Swinbank} {et~al}\mbox{.}(2015){Swinbank}, {Dye}, {Nightingale},
  {Furlanetto}, {Smail}, {Cooray}, {Dannerbauer}, {Dunne}, {Eales}, {Gavazzi},
  {Hunter}, {Ivison}, {Negrello}, {Oteo-Gomez}, {Smit}, {van der Werf}, \&
  {Vlahakis}}]{Swinbank2015ApJ...806L..17S}
{Swinbank} A.~M. {et~al.}, 2015, \apjl, 806, L17

\bibitem[{{Swinbank} {et~al}\mbox{.}(2012){Swinbank}, {Karim}, {Smail},
  {Hodge}, {Walter}, {Bertoldi}, {Biggs}, {de Breuck}, {Chapman}, {Coppin},
  {Cox}, {Danielson}, {Dannerbauer}, {Ivison}, {Greve}, {Knudsen}, {Menten},
  {Simpson}, {Schinnerer}, {Wardlow}, {Wei{\ss}}, \& {van der
  Werf}}]{Swinbank2012MNRAS.427.1066S}
{Swinbank} A.~M. {et~al.}, 2012, \mnras, 427, 1066

\bibitem[{{Swinbank} {et~al}\mbox{.}(2014){Swinbank}, {Simpson}, {Smail},
  {Harrison}, {Hodge}, {Karim}, {Walter}, {Alexander}, {Brandt}, {de Breuck},
  {da Cunha}, {Chapman}, {Coppin}, {Danielson}, {Dannerbauer}, {Decarli},
  {Greve}, {Ivison}, {Knudsen}, {Lagos}, {Schinnerer}, {Thomson}, {Wardlow},
  {Wei{\ss}}, \& {van der Werf}}]{Swinbank2014MNRAS.438.1267S}
{Swinbank} A.~M. {et~al.}, 2014, \mnras, 438, 1267

\bibitem[{{Symeonidis} {et~al}\mbox{.}(2013){Symeonidis}, {Vaccari}, {Berta},
  {Page}, {Lutz}, {Arumugam}, {Aussel}, {Bock}, {Boselli}, {Buat}, {Capak},
  {Clements}, {Conley}, {Conversi}, {Cooray}, {Dowell}, {Farrah},
  {Franceschini}, {Giovannoli}, {Glenn}, {Griffin}, {Hatziminaoglou}, {Hwang},
  {Ibar}, {Ilbert}, {Ivison}, {Floc'h}, {Lilly}, {Kartaltepe}, {Magnelli},
  {Magdis}, {Marchetti}, {Nguyen}, {Nordon}, {O'Halloran}, {Oliver}, {Omont},
  {Papageorgiou}, {Patel}, {Pearson}, {P{\'e}rez-Fournon}, {Pohlen}, {Popesso},
  {Pozzi}, {Rigopoulou}, {Riguccini}, {Rosario}, {Roseboom}, {Rowan-Robinson},
  {Salvato}, {Schulz}, {Scott}, {Seymour}, {Shupe}, {Smith}, {Valtchanov},
  {Wang}, {Xu}, {Zemcov}, \& {Wuyts}}]{Symeonidis2013MNRAS.431.2317S}
{Symeonidis} M. {et~al.}, 2013, \mnras, 431, 2317

\bibitem[{{Tacconi} {et~al}\mbox{.}(2008){Tacconi}, {Genzel}, {Smail}, {Neri},
  {Chapman}, {Ivison}, {Blain}, {Cox}, {Omont}, {Bertoldi}, {Greve},
  {F{\"o}rster Schreiber}, {Genel}, {Lutz}, {Swinbank}, {Shapley}, {Erb},
  {Cimatti}, {Daddi}, \& {Baker}}]{Tacconi2008ApJ...680..246T}
{Tacconi} L.~J. {et~al.}, 2008, \apj, 680, 246

\bibitem[{{Tacconi} {et~al}\mbox{.}(2006){Tacconi}, {Neri}, {Chapman},
  {Genzel}, {Smail}, {Ivison}, {Bertoldi}, {Blain}, {Cox}, {Greve}, \&
  {Omont}}]{Tacconi2006ApJ...640..228T}
{Tacconi} L.~J. {et~al.}, 2006, \apj, 640, 228

\bibitem[{{Thomas} {et~al}\mbox{.}(2005){Thomas}, {Maraston}, {Bender}, \&
  {Mendes de Oliveira}}]{Thomas2005ApJ...621..673T}
{Thomas} D., {Maraston} C., {Bender} R., {Mendes de Oliveira} C., 2005, \apj,
  621, 673

\bibitem[{{Thomas} {et~al}\mbox{.}(2010){Thomas}, {Maraston}, {Schawinski},
  {Sarzi}, \& {Silk}}]{Thomas2010MNRAS.404.1775T}
{Thomas} D., {Maraston} C., {Schawinski} K., {Sarzi} M., {Silk} J., 2010,
  \mnras, 404, 1775

\bibitem[{{Toomre}(1964)}]{Toomre1964ApJ...139.1217T}
{Toomre} A., 1964, \apj, 139, 1217

\bibitem[{{Tozzi} {et~al}\mbox{.}(2015){Tozzi}, {Santos}, {Jee}, {Fassbender},
  {Rosati}, {Nastasi}, {Forman}, {Sartoris}, {Borgani}, {Boehringer},
  {Altieri}, {Pratt}, {Nonino}, \& {Jones}}]{Tozzi2015ApJ...799...93T}
{Tozzi} P. {et~al.}, 2015, \apj, 799, 93

\bibitem[{{Walter} {et~al}\mbox{.}(2012){Walter}, {Decarli}, {Carilli},
  {Bertoldi}, {Cox}, {da Cunha}, {Daddi}, {Dickinson}, {Downes}, {Elbaz},
  {Ellis}, {Hodge}, {Neri}, {Riechers}, {Weiss}, {Bell}, {Dannerbauer},
  {Krips}, {Krumholz}, {Lentati}, {Maiolino}, {Menten}, {Rix}, {Robertson},
  {Spinrad}, {Stark}, \& {Stern}}]{Walter2012Natur.486..233W}
{Walter} F. {et~al.}, 2012, \nat, 486, 233

\bibitem[{{Walter} {et~al}\mbox{.}(2009){Walter}, {Riechers}, {Cox}, {Neri},
  {Carilli}, {Bertoldi}, {Weiss}, \& {Maiolino}}]{Walter2009Natur.457..699W}
{Walter} F., {Riechers} D., {Cox} P., {Neri} R., {Carilli} C., {Bertoldi} F.,
  {Weiss} A., {Maiolino} R., 2009, \nat, 457, 699

\bibitem[{{Wei{\ss}} {et~al}\mbox{.}(2009){Wei{\ss}}, {Kov{\'a}cs}, {Coppin},
  {Greve}, {Walter}, {Smail}, {Dunlop}, {Knudsen}, {Alexander}, {Bertoldi},
  {Brandt}, {Chapman}, {Cox}, {Dannerbauer}, {De Breuck}, {Gawiser}, {Ivison},
  {Lutz}, {Menten}, {Koekemoer}, {Kreysa}, {Kurczynski}, {Rix}, {Schinnerer},
  \& {van der Werf}}]{Weiss2009ApJ...707.1201W}
{Wei{\ss}} A. {et~al.}, 2009, \apj, 707, 1201

\bibitem[{{Wolfire} {et~al}\mbox{.}(2010){Wolfire}, {Hollenbach}, \&
  {McKee}}]{Wolfire2010ApJ...716.1191W}
{Wolfire} M.~G., {Hollenbach} D., {McKee} C.~F., 2010, \apj, 716, 1191

\bibitem[{{Zhu} {et~al}\mbox{.}(2007){Zhu}, {Gao}, {Seaquist}, \&
  {Dunne}}]{Zhu2007AJ....134..118Z}
{Zhu} M., {Gao} Y., {Seaquist} E.~R., {Dunne} L., 2007, \aj, 134, 118

\bibitem[{{Zirm} {et~al}\mbox{.}(2008){Zirm}, {Stanford}, {Postman},
  {Overzier}, {Blakeslee}, {Rosati}, {Kurk}, {Pentericci}, {Venemans}, {Miley},
  {R{\"o}ttgering}, {Franx}, {van der Wel}, {Demarco}, \& {van
  Breugel}}]{Zirm2008ApJ...680..224Z}
{Zirm} A.~W. {et~al.}, 2008, \apj, 680, 224

\end{thebibliography}

\end{document}